\documentclass{aa}  

\usepackage{graphicx}
\usepackage{wasysym}
\usepackage{amssymb}
\usepackage{subfigure}
\usepackage{txfonts}
\usepackage[]{hyperref}
\usepackage{breakurl}
\usepackage{upgreek}

\begin{document}

   \title{Spectroscopic study of solar twins and analogues
    \thanks{Based on observations made with ESO
    Telescopes at the La Silla Observatory under programme ID 077.D-0525 and 090.D-0133.}
    \thanks{Table 1 and the full Table 5 are available in electronic form at the CDS via anonymous ftp to cdsarc.u-strasbg.fr (130.79.128.5) or via http://cdsweb.u-strasbg.fr/cgi-bin/qcat?J/A+A/}}

   \subtitle{}

   \author{Juliet Datson\inst{1}\fnmsep\thanks{\email{juliet.datson@gmail.com}}
          \and
          Chris Flynn\inst{2,3,4}
          \and
          Laura Portinari\inst{1}
          }

   \institute{Tuorla Observatory, Department of Physics and Astronomy, University of Turku, V\"ais\"al\"antie 20, FI-21500 Piikki\"o, Finland
        \and
                Centre for Astrophysics and Supercomputing, Swinburne University of Technology, VIC 3122 Australia
        \and
                Finnish Centre for Astronomy with ESO, University of Turku, FI-21500, Piikkio, Finland
        \and
                Department of Physics and Astronomy, University of Sydney, NSW 2006 Australia
                        }

   \date{Received 00 XXXX 2014 / Accepted 00 XXXX 2014}

 
  \abstract
   {Many large stellar surveys have been and are still being carried out, providing huge amounts of data, for which stellar physical parameters will be derived. Solar twins and analogues provide a means to test the calibration of these stellar catalogues
because the Sun is the best-studied star and provides precise fundamental parameters. Solar twins should be centred on the solar values.} 
   {This spectroscopic study of solar analogues selected from the Geneva-Copenhagen Survey (GCS) at a resolution of 48,000 provides effective temperatures and metallicities for these stars. We test whether our spectroscopic parameters, as well as the previous photometric calibrations, are properly centred on the Sun. In addition, we search for more solar twins in our sample.}
   {The methods used in this work are based on literature methods for solar twin searches and on methods we developed in previous work to distinguish the metallicity-temperature degeneracies in the differential comparison of spectra of solar analogues versus a reference solar reflection spectrum.}
   {We derive spectroscopic parameters for 148 solar analogues (about 70 are new entries to the literature) and verify with a-posteriori differential tests that our values are well-centred on the solar values. We use our dataset to assess the two alternative calibrations of the GCS parameters; our methods favour the latest revision. We show that the choice of spectral line list or the choice of asteroid or time of observation does not affect the results. We also identify seven solar twins in our sample, three of which are published here for the first time.}
   {Our methods provide an independent means to differentially test the calibration of stellar catalogues around the values of a well-known benchmark star, which makes our work interesting
for calibration tests of upcoming Galactic surveys.}


   \maketitle

\section{Introduction}

Large stellar surveys of the Milky Way are becoming increasingly important in astronomy. Whether it is data from past missions such as the Two Micron All-Sky Survey \citep[2MASS, ][]{Sk95}, HIPPARCOS \citep{Pe97,Le07}, the Sloan Digital Sky Survey \citep[SDSS, ][]{Yo00}, the Geneva-Copenhagen-Survey \citep[GCS, ][]{No04,Ho07,Ho09}, or the RAdial Velocity Experiment \citep[RAVE, ][]{St03}, or from current and future surveys such as HERMES/GALAH \citep{Fr10}, the Gaia-ESO Survey \citep[GES, ][]{Gi12} or Gaia \citep{Mu03} -- they all have one thing in common: the wish to better understand the Milky Way and through it, galaxy evolution in general.

To address questions about and problems of the chemical evolution of galaxies, the stellar mass and luminosity functions, the star formation history of the Milky Way, dynamical evolution within the disc, radial migration of stars, etc., it is necessary to determine precise and accurate physical parameters for the observed stars. The methods used to provide these vary from survey to survey because they are based on different wavelength regimes and types of data (photometry or spectroscopy), which makes it particularly important to ensure that these catalogues are properly calibrated.

In previous work \citet[D12 and D14 hereafter]{Da12,Da14} have shown that by using a sample of photometrically chosen solar analogue stars and applying spectroscopic methods that are based on solar twin searches, it is possible to independently test the calibration of a catalogue around the values of a benchmark star by a purely differential analysis. This star needs to have well-known parameters and be included in the observations. So far, in D12 and D14 this was achieved by using the Sun, that
is, a solar reflection spectrum of an asteroid like Ceres, as a reference solar spectrum.

By doubling the sample size of D12, we improve the accuracy of our analysis and test whether the results we obtained in that paper change with higher number of targets, meaning that we test whether the previously found offsets in the GCS-III changed. The larger sample size also allows us to search for more solar twins in the data and inspect the reanalysis of the Geneva-Copenhagen Survey \citep[C11, ][]{Ca11} by dividing the sample with method of stellar parameter determination (infrared-flux method versus colour calibrations) to test the two calibrations separately.

{\sc moog} \citep{Sn73} is a very widely used tool to determine spectroscopic stellar parameters, therefore we also used it here to determine the characteristics of the sample stars. One source of error and discussion is the choice of spectral lines used for the elemental abundance determination. While the chosen lines are expected to be weak, unblended, and isolated lines on the linear part of the curve of growth, the exact list always varies \citep[e.g.][]{Me12,Po14}. The choice of lines matters, and a careful selection of lines is key to a precise analysis \citep{So14, Jo14}. For this reason, we chose four different line lists here, all selected by the respective authors to fulfil the above-mentioned requirements, to test whether our conclusions on the solar zero point of a given catalogue calibration and solar twin determination are reliable or depend on the line list.

Another source of systematic error could be the specific choice of the reflected solar spectrum. Typically, one or several of the brightest asteroids and/or Jovian moons are used \citep[e.g.][]{Ra14a, Ni14, Me14}. We took the opportunity of taking a total of 15 spectra of the asteroids Ceres (8) and Vesta (7) at several different times during our three-night run to test for significant differences in using one asteroid or the other.

The structure of the paper is as follows: In Sect. 2 we present the data. Section 3 provides new spectroscopic parameters for our sample stars and compares them with literature values. In Sect. 4 we search for more solar twins in our new sample, and in Sect. 5 we reassess the calibration of GCS-III versus its reanalysis (C11), including a possible dependency of our analysis on the chosen line list. Section 6 is a test on asteroid spectra stability, and we conclude in Sect. 7.

\section{Observations and data}

Our data consist of two samples of spectra, taken with the Fiber-fed Extended Range Optical Spectrograph (FEROS) \citep{Ka99} on the MPG/ESO 2.2m telescope on La Silla, Chile. FEROS is a high-resolution ($R = 48,000$) echelle spectrograph that covers a spectral range of $\lambda = 3500 - 9200\AA$. We used the same sample as in D12 (FE12 sample hereafter) and added a new FEROS sample (FE14 sample hereafter). 

\subsection{FE12 sample}

The solar analogues were initially chosen from the Geneva-Copenhagen
Survey releases I and III \citep{No04, Ho09} by bracketing the solar $(b-y)$ colour, absolute visual magnitude $M_{V}$ , and photometric metallicity, for which we adopted the solar values of $(b-y)_\odot = 0.403$ \citep{Ho06}, $M_{V} = 4.83$ \citep{Al76} and [Fe/H] $=0.0$ (by definition). This resulted in 338 stars, of which we selected 70 that were observable in July and August 2006 with exposure times of $200 - 950$~s for a signal-to-noise (S/N) ratio of $\sim$150. Additionally, we searched the FEROS archive for a few other known solar analogues to add to our sample and found another 75 stars, whose spectra were taken between 2003 and 2008 and also achieved an S/N ratio of the aimed $\sim$150. This gave us spectra of 145 stars in the FE12 sample before data reduction and analysis. For the solar comparison we also added a spectrum of the asteroid Ceres with a similar S/N ratio. For details, see D12.

\subsection{FE14 sample}

For the FE14 sample we chose the same selection criteria with a minor adjustment. As we have shown in D12 and D14, there seems to be an offset of $-0.10$~dex in metallicity for solar-type stars in the GCS, therefore we expanded the metallicity criterium to $-0.25 < [\mathrm{Fe/H}] < 0.15$, allowing more metal-poor stars into the sample to better bracket the solar values. Taking into account the observability of the targets during the run and choosing the brightest targets, this resulted in 88 stars, 70 of which were observed. There is no overlap with the FE12 sample.

The data for this sample were taken between $21^{\mathrm{}} - 24^{\mathrm{}}$ December 2012 with the FEROS instrument on the MPG/ESO 2.2m telescope on La Silla. Exposure times ranged from 90s to 1200s, aiming for an S/N ratio of 200. In addition to the target stars, several spectra of the asteroids Ceres and Vesta were included as solar comparisons. Although FEROS is very stable instrument, there were some inconsistencies in the data reduction process, which introduced some noise into the extracted spectra, effectively reducing the quality of the spectra.

Both samples include repeated spectra of some stars for internal consistency checks and error estimates.

\subsection{Data reduction}

The data were reduced in the same way as in D12 by using the FEROS pipeline, which automatically uses a set of nightly calibration frames to flat-field, bias subtract, sky-subtract, and wavelength-calibrate the science spectra and delivers 1D spectra for the subsequent analysis. It uses ThAr+Ne arcs for the wavelength calibration and rebins the spectra to a linear 0.03\AA\ resolution over the whole wavelength range.

The resulting spectra also show significant wiggles and jumps in the continuum level, however, therefore we additionally flattened
them by fitting piecewise 10\AA\ sections of the spectrum. We used the same approach in D12. This resulted in flat spectra in the regions of isolated lines, which we used for our further analysis.

Unfortunately, not all spectra could be flattened and some had to be discarded. In the end, we had 161 spectra of 148 stars to use in our samples: 96 in the FE12 and 65 in the FE14 sample. 

\begin{figure}
  \centering 
  \includegraphics[scale=0.4]{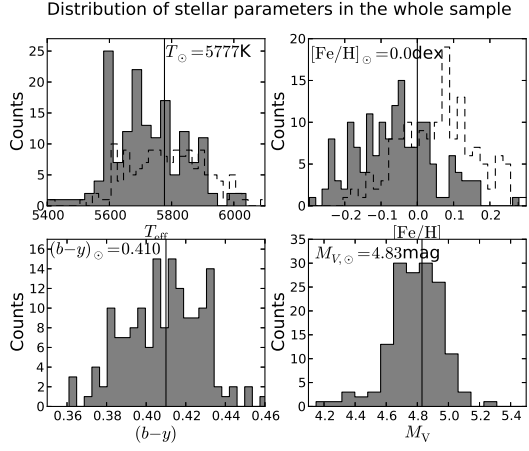}
  \caption{Distribution of stellar parameters in the combined FEplus sample. The top two panels show values from the GCS in the shaded histograms and the values from our own spectroscopic determinations as the dashed histograms. The bottom two panels show values from the GCS for $(b-y)$ and HIPPARCOS for $M_{V}$. Vertical black lines show our adopted solar values of all four parameters. We quote a solar $(b-y)$ colour that we found in previous work (D12 and D14) and not the lower value by \citet{Ho06}.}
  \label{param}
\end{figure}

In Fig.~\ref{param} we show the distribution of the stellar parameters in our combined sample (FEplus hereafter), as taken from the GCS catalogue. The stars were chosen to best cover the pre-selected parameter range and are not expected to be representative of the local volume.

\section{Spectroscopic temperatures and metallicities}
\label{specparam}

To determine our own values for the effective temperatures and metallicities in the FEplus sample stars, we used the local thermodynamic equilibrium (LTE) spectrum analysis code {\sc moog} by \citet{Sn73} with the 2012 version of the MARCS models \citep{Gu08}. Furthermore, we used the {\sc smh} (spectroscopy made hard) wrapper code by \citet{Cs14}, which implements the measurement of the equivalent widths of the spectral lines (automatically and/or by hand) and uses {\sc moog} to determine the stellar parameters through the standard process of iteration on the ionisation and excitation equilibria.

For this analysis we used the line list by \citet{Bi12}, optimised for our wavelength range and S/N, consisting of 151 spectral lines of nine different neutral elements and one ionised species. They are Na\thinspace I, Al\thinspace I, Si\thinspace I, Ca\thinspace I, Ti\thinspace I, Cr\thinspace I, Fe\thinspace I+II, Ni\thinspace I, and Zn\thinspace I. Of the 151 spectral lines, 80 were from neutral and 7 from singly ionised iron, which is the only element necessary for the stellar parameter determination. 

For both samples (FE12 and FE14) we had one Ceres spectrum each, which worked as a reference solar spectrum for that specific sample. We used these solar reference spectra to determine a zero point for the analysis, that is, the {\sc smh} code gave us formal results for $T_{\mathrm{eff}}$, [Fe/H] and log$g$, which showed an offset to the real solar values. For FE12 we found $T_{\mathrm{eff}} = 5815$~K, [Fe/H] = 0.08~dex and log$g$ = 4.33; for FE14 they were $T_{\mathrm{eff}} = 5813$~K, [Fe/H] = 0.15~dex and log$g$ = 4.29. Compared to the known solar values, the formal absolute accuracy is clearly worse than the precision, especially for [Fe/H] (see below); but as we are analysing solar analogues, our results are to be considered in terms of difference from the Sun. Therefore we applied a systematic correction to all our results, so that the values from the Ceres spectra are scaled to be exactly solar in each respective sample (FE12 and FE14). The final results, after correction, can be seen in Table \ref{newvalues}. We determined the precision of the individual measurements through comparing stellar parameters of the same object and found typical errors of $\sigma (T_{\mathrm{eff}}) = 40$~K, $\sigma$([Fe/H]) = 0.03~dex and $\sigma (\log g)$ = 0.07.

\begin{table*}
\caption{New spectroscopic parameters for the stars from the FEplus sample derived in this work; twins are highlighted in boldface. This table is also available at the CDS.}
\label{newvalues}
\small
\centering
\scalebox{0.9}{
\begin{tabular}{ccccc|ccccc}
\hline
HD number & Sample & $T_{\mathrm{eff}}$ [K] & [Fe/H] [dex] & log g [dex] & HD number & Sample & $T_{\mathrm{eff}}$ [K] & [Fe/H] [dex] & log g [dex]\\
\hline
1835 & FE12 & 5849 & \hphantom{$-$}0.22 & 4.61 & 64474 & FE14 & 5656 & $-$0.02 & 4.25\\
4392 & FE14 & 5656 & \hphantom{$-$}0.05 & 4.34 & 64942 & FE12 & 5839 & \hphantom{$-$}0.04 & 4.45\\
4428 & FE14 & 5740 & $-$0.19 & 4.49 & 65243 & FE14 & 5798 & $-$0.10 & 4.39\\
6715 & FE14 & 5621 & $-$0.19 & 4.44 & 66653 & FE14 & 5809 & \hphantom{$-$}0.09 & 4.42\\
7570 & FE12 & 6146 & \hphantom{$-$}0.18 & 4.35 & \bf{67010} & \bf{FE14} & \bf{5613} & \bf{$-$0.13} & \bf{4.48}\\
7678 & FE12 & 5844 & \hphantom{$-$}0.14 & 4.52 & 68168 & FE14 & 5638 & \hphantom{$-$}0.02 & 4.22\\
7727 & FE12 & 6089 & \hphantom{$-$}0.07 & 4.32 & 70642 & FE12 & 5590 & \hphantom{$-$}0.20 & 4.44\\
8076 & FE12 & 5893 & \hphantom{$-$}0.05 & 4.43 & 73350 & FE14 & 5874 & \hphantom{$-$}0.13 & 4.56\\
8129 & FE14 & 5597 & $-$0.04 & 4.49 & 75302 & FE12 & 5727 & \hphantom{$-$}0.14 & 4.57\\
9562 & FE12 & 5808 & \hphantom{$-$}0.14 & 3.98 & 75767 & FE14 & 5859 & $-$0.01 & 4.45\\
9782 & FE12 & 6009 & \hphantom{$-$}0.08 & 4.39 & 75880 & FE14 & 5664 & \hphantom{$-$}0.14 & 4.27\\
9986 & FE12 & 5828 & \hphantom{$-$}0.11 & 4.49 & 76151 & FE12 & 5752 & \hphantom{$-$}0.12 & 4.51\\
10180 & FE12 & 5891 & \hphantom{$-$}0.07 & 4.36 & 76332 & FE14 & 5827 & $-$0.06 & 4.34\\
10226 & FE12 & 5989 & \hphantom{$-$}0.12 & 4.47 & \bf{76440} & \bf{FE14} & \bf{5781} & \bf{$-$0.05} & \bf{4.34}\\
10700 & FE12 & 5425 &$-$0.52 & 4.90 & 76780 & FE14 & 5745 & \hphantom{$-$}0.11 & 4.40\\
11131 & FE12 & 5804 & $-$0.09 & 4.53 & 77461 & FE14 & 5835 & $-$0.02 & 4.49\\
12264 & FE12 & 5819 & \hphantom{$-$}0.05 & 4.47 & 78317 & FE12 & 5848 & \hphantom{$-$}0.05 & 4.43\\
13043 & FE12 & 5870 & \hphantom{$-$}0.06 & 4.23 & 78538 & FE12 & 5800 & $-$0.02 & 4.59\\
13382 & FE14 & 5799 & \hphantom{$-$}0.19 & 4.51 & \bf{78660} & \bf{FE12} & \bf{5776} & \bf{\hphantom{$-$}0.06} & \bf{4.55}\\
13386 & FE12 & 5345 & \hphantom{$-$}0.23 & 4.68 & 81659 & FE12 & 5658 & \hphantom{$-$}0.23 & 4.52\\
13724 & FE12 & 5809 & \hphantom{$-$}0.25 & 4.54 & 86226 & FE12 & 5934 & \hphantom{$-$}0.01 & 4.37\\
13825 & FE14 & 5622 & \hphantom{$-$}0.17 & 4.37 & 87359 & FE12 & 5679 & \hphantom{$-$}0.09 & 4.54\\
15632 & FE14 & 5762 & \hphantom{$-$}0.03 & 4.51 & 88072 & FE12 & 5759 & \hphantom{$-$}0.08 & 4.53\\
16417 & FE12 & 5766 & \hphantom{$-$}0.09 & 4.16 & 89055 & FE14 & 5745 & $-$0.19 & 4.26\\
18330 & FE12 & 5989 & $-$0.03 & 4.52 & 89454 & FE12 & 5690 & \hphantom{$-$}0.10 & 4.48\\
19518 & FE12 & 5781 & $-$0.07 & 4.60 & 90936 & FE12 & 5918 & \hphantom{$-$}0.07 & 4.43\\
19617 & FE14 & 5727 & \hphantom{$-$}0.15 & 4.38 & 91489 & FE14 & 5884 & $-$0.01 & 4.48\\
19632 & FE12 & 5748 & \hphantom{$-$}0.17 & 4.59 & 93215 & FE14 & 5558 & \hphantom{$-$}0.04 & 4.09\\
20201 & FE12 & 5994 & \hphantom{$-$}0.10 & 4.44 & 93489 & FE12 & 5904 & \hphantom{$-$}0.04 & 4.41\\
20527 & FE14 & 5615 & \hphantom{$-$}0.11 & 4.45 & 94151 & FE12 & 5608 & \hphantom{$-$}0.07 & 4.59\\
22623 & FE14 & 5520 & $-$0.04 & 4.45 & \bf{97356} & \bf{FE12} & \bf{5805} & \bf{\hphantom{$-$}0.02} & \bf{4.35}\\
24293 & FE14 & 5674 & $-$0.11 & 4.32 & 98222 & FE12 & 5666 & \hphantom{$-$}0.08 & 4.35\\
24552 & FE12 & 5825 & $-$0.06 & 4.44 & 98649 & FE12 & 5687 & $-$0.06 & 4.56\\
\bf{25680} & \bf{FE14} & \bf{5905} & \bf{\hphantom{$-$}0.08} & \bf{4.57} & 98764 & FE12 & 5951 & \hphantom{$-$}0.15 & 4.68\\
25710 & FE14 & 5994 & $-$0.09 & 4.42 & 99610 & FE14 & 5656 & \hphantom{$-$}0.16 & 4.32\\
25874 & FE14 & 5715 & $-$0.04 & 4.32 & 101530 & FE14 & 5839 & $-$0.19 & 4.41\\
25926 & FE14 & 5915 & $-$0.01 & 4.38 & 106252 & FE12 & 5927 & $-$0.02 & 4.38\\
26151 & FE12 & 5394 & \hphantom{$-$}0.24 & 4.68 & 108523 & FE12 & 5620 & \hphantom{$-$}0.05 & 4.58\\
26767 & FE14 & 5792 & \hphantom{$-$}0.08 & 4.30 & 110668 & FE12 & 5819 & \hphantom{$-$}0.19 & 4.42\\
27685 & FE14 & 5705 & \hphantom{$-$}0.07 & 4.42 & \bf{117860} & \bf{FE12} & \bf{5948} & \bf{\hphantom{$-$}0.13} & \bf{4.62}\\
28068 & FE14 & 5761 & \hphantom{$-$}0.07 & 4.32 & 119856 & FE12 & 5894 & $-$0.14 & 4.58\\
28099 & FE14 & 5718 & \hphantom{$-$}0.06 & 4.38 & 122973 & FE12 & 5982 & \hphantom{$-$}0.12 & 4.51\\
29150 & FE14 & 5755 & $-$0.02 & 4.48 & 125612 & FE12 & 5874 & \hphantom{$-$}0.22 & 4.51\\
29161 & FE14 & 5890 & $-$0.03 & 4.59 & 126053 & FE12 & 5655 & $-$0.40 & 4.48\\
29461 & FE14 & 5766 & \hphantom{$-$}0.16 & 4.38 & \bf{126525} & \bf{FE12} & \bf{5628} & \bf{$-$0.02} & \bf{4.52}\\
\bf{29601} & \bf{FE14} & \bf{5608} & \bf{$-$0.10} & \bf{4.24} & 134664 & FE12 & 5884 & \hphantom{$-$}0.13 & 4.43\\
30246 & FE14 & 5723 & \hphantom{$-$}0.08 & 4.52 & \bf{138573} & \bf{FE12} & \bf{5777} & \bf{\hphantom{$-$}0.00} & \bf{4.46}\\
30306 & FE12 & 5557 & \hphantom{$-$}0.17 & 4.53 & 142072 & FE12 & 5761 & \hphantom{$-$}0.14 & 4.38\\
30495 & FE12 & 5769 & $-$0.04 & 4.45 & \bf{142415} & \bf{FE12} & \bf{5904} & \bf{\hphantom{$-$}0.08} & \bf{4.40}\\
30774 & FE14 & 5665 & \hphantom{$-$}0.02 & 4.34 & 145518 & FE12 & 5900 & $-$0.06 & 4.49\\
31222 & FE14 & 5659 & \hphantom{$-$}0.12 & 4.50 & 145666 & FE12 & 5915 & \hphantom{$-$}0.01 & 4.38\\
31622 & FE14 & 5823 & $-$0.17 & 4.52 & 146070 & FE12 & 5821 & $-$0.09 & 4.43\\
32963 & FE12 & 5710 & \hphantom{$-$}0.07 & 4.42 & \bf{146233} & \bf{FE12} & \bf{5819} & \bf{\hphantom{$-$}0.08} & \bf{4.47}\\
33866 & FE14 & 5836 & $-$0.03 & 4.45 & \bf{147513} & \bf{FE12} & \bf{5885} & \bf{\hphantom{$-$}0.04} & \bf{4.53}\\
33873 & FE14 & 5658 & \hphantom{$-$}0.19 & 4.37 & 152322 & FE12 & 5949 & \hphantom{$-$}0.02 & 4.57\\
34239 & FE14 & 5932 & \hphantom{$-$}0.04 & 4.40 & 155114 & FE12 & 5791 & $-$0.07 & 4.56\\
34386 & FE14 & 5689 & \hphantom{$-$}0.07 & 4.25 & 155968 & FE12 & 5720 & \hphantom{$-$}0.16 & 4.47\\
34599 & FE14 & 5834 & \hphantom{$-$}0.08 & 4.40 & \bf{163441} & \bf{FE12} & \bf{5795} & \bf{\hphantom{$-$}0.09} & \bf{4.43}\\
\bf{35769} & \bf{FE14} & \bf{5631} & \bf{$-$0.13} & \bf{4.34} & 168746 & FE12 & 5572 & $-$0.07 & 4.44\\
36152 & FE12 & 5755 & \hphantom{$-$}0.08 & 4.50 & \bf{173071} & \bf{FE12} & \bf{6044} & \bf{\hphantom{$-$}0.25} & \bf{4.49}\\
36553 & FE12 & 6017 & \hphantom{$-$}0.32 & 3.81 & 183505 & FE12 & 5716 & \hphantom{$-$}0.14 & 4.52\\
37773 & FE14 & 5618 & $-$0.04 & 4.25 & 189931 & FE12 & 5925 & \hphantom{$-$}0.03 & 4.53\\
\bf{39649} & \bf{FE14} & \bf{5740} & \bf{\hphantom{$-$}0.01} & \bf{4.40} & 192417 & FE12 & 5745 & \hphantom{$-$}0.05 & 4.56\\
39833 & FE12 & 5821 & \hphantom{$-$}0.16 & 4.38 & 202628 & FE12 & 5848 & \hphantom{$-$}0.02 & 4.60\\
\bf{41708} & \bf{FE14} & \bf{5928} & \bf{\hphantom{$-$}0.08} & \bf{4.45} & 207043 & FE12 & 5731 & \hphantom{$-$}0.03 & 4.46\\
43745 & FE12 & 5958 & $-$0.05 & 3.84 & 209262 & FE12 & 5760 & \hphantom{$-$}0.13 & 4.40\\
44665 & FE12 & 5694 & \hphantom{$-$}0.00 & 4.44 & 212708 & FE12 & 5685 & \hphantom{$-$}0.29 & 4.48\\
46090 & FE14 & 5778 & \hphantom{$-$}0.01 & 4.46 & 213199 & FE12 & 5908 & $-$0.04 & 4.41\\
47186 & FE12 & 5627 & \hphantom{$-$}0.23 & 4.43 & 214954 & FE12 & 5727 & \hphantom{$-$}0.18 & 4.50\\
48969 & FE14 & 5685 & $-$0.07 & 4.53 & 215657 & FE12 & 5989 & \hphantom{$-$}0.07 & 4.42\\
51219 & FE14 & 5608 & \hphantom{$-$}0.01 & 4.34 & 218205 & FE12 & 5945 & \hphantom{$-$}0.12 & 4.55\\
55693 & FE12 & 5855 & \hphantom{$-$}0.24 & 4.40 & 221343 & FE12 & 5822 & \hphantom{$-$}0.12 & 4.58\\
58895 & FE12 & 5690 & \hphantom{$-$}0.24 & 4.02 & 222669 & FE12 & 5877 & \hphantom{$-$}0.09 & 4.46\\
59967 & FE14 & 5860 & $-$0.03 & 4.58 & 225299 & FE12 & 5754 & \hphantom{$-$}0.23 & 4.56\\
\hline
\end{tabular}
}
\end{table*}

\subsection{Testing the calibration}
\label{deglin}

\begin{figure*}
\centering
\subfigure{
  \includegraphics[scale=0.35]{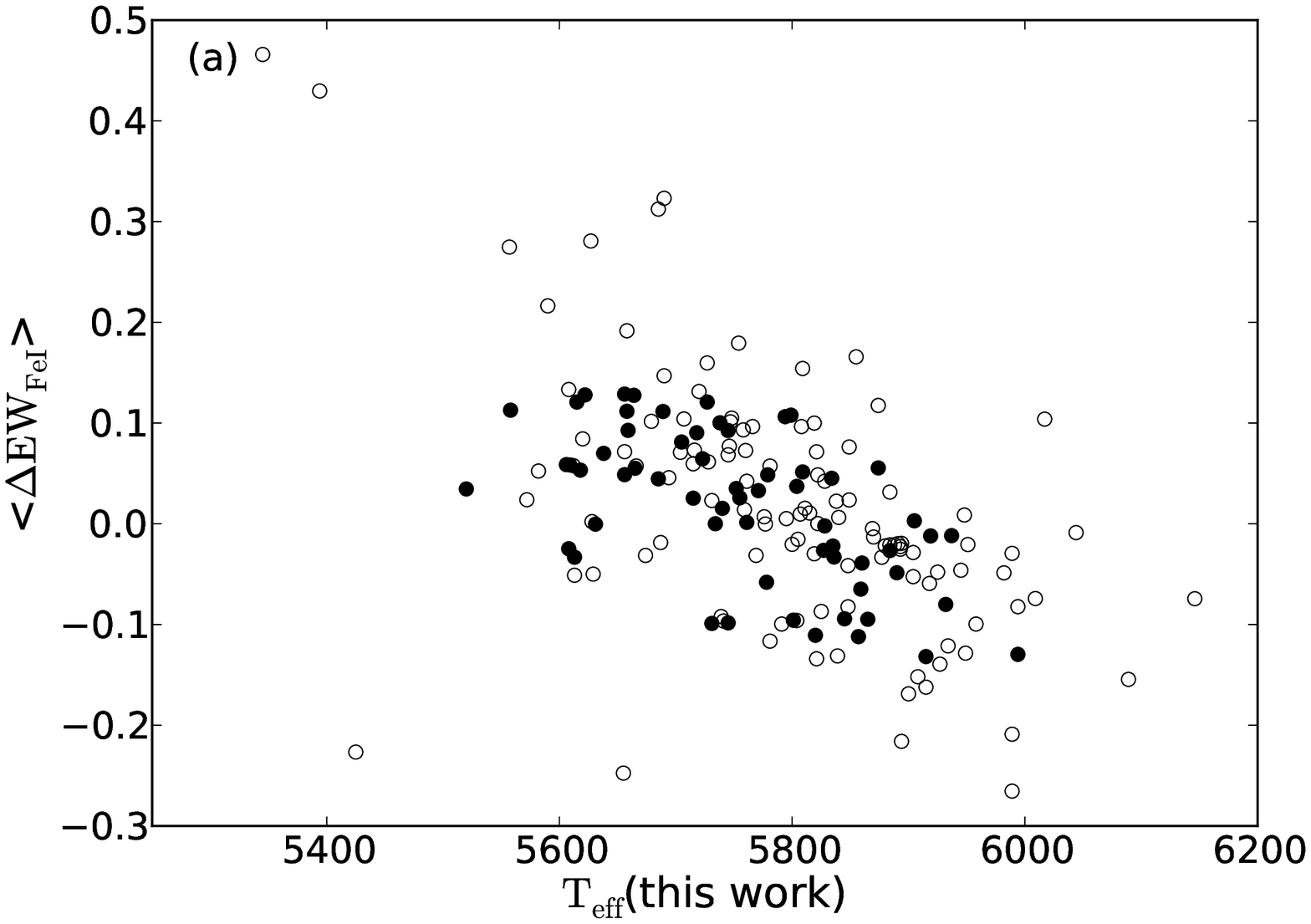}
\label{fig:subfig5}
}
\subfigure{
  \includegraphics[scale=0.35]{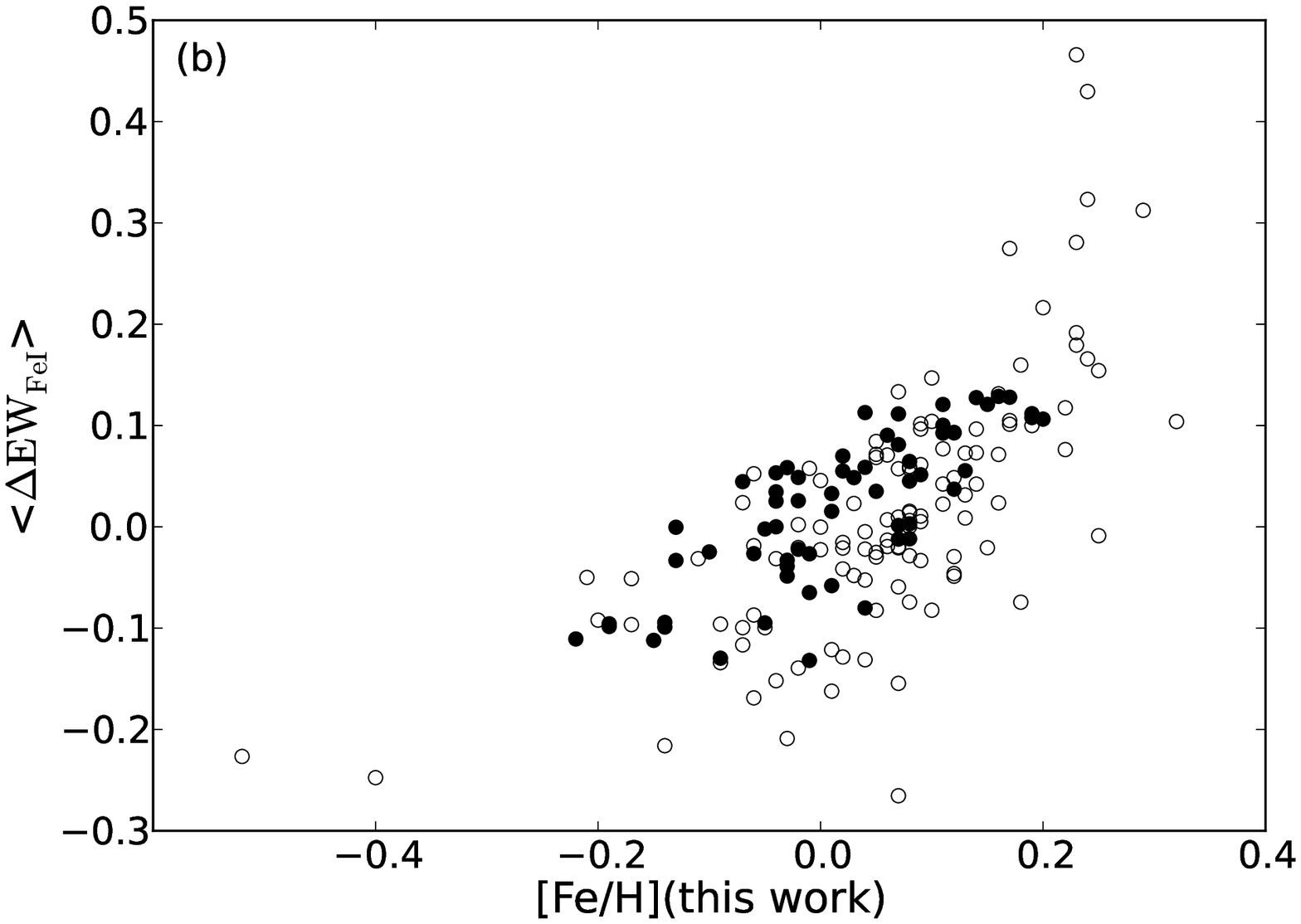}
\label{fig:subfig6}
}
\subfigure{
  \includegraphics[scale=0.35]{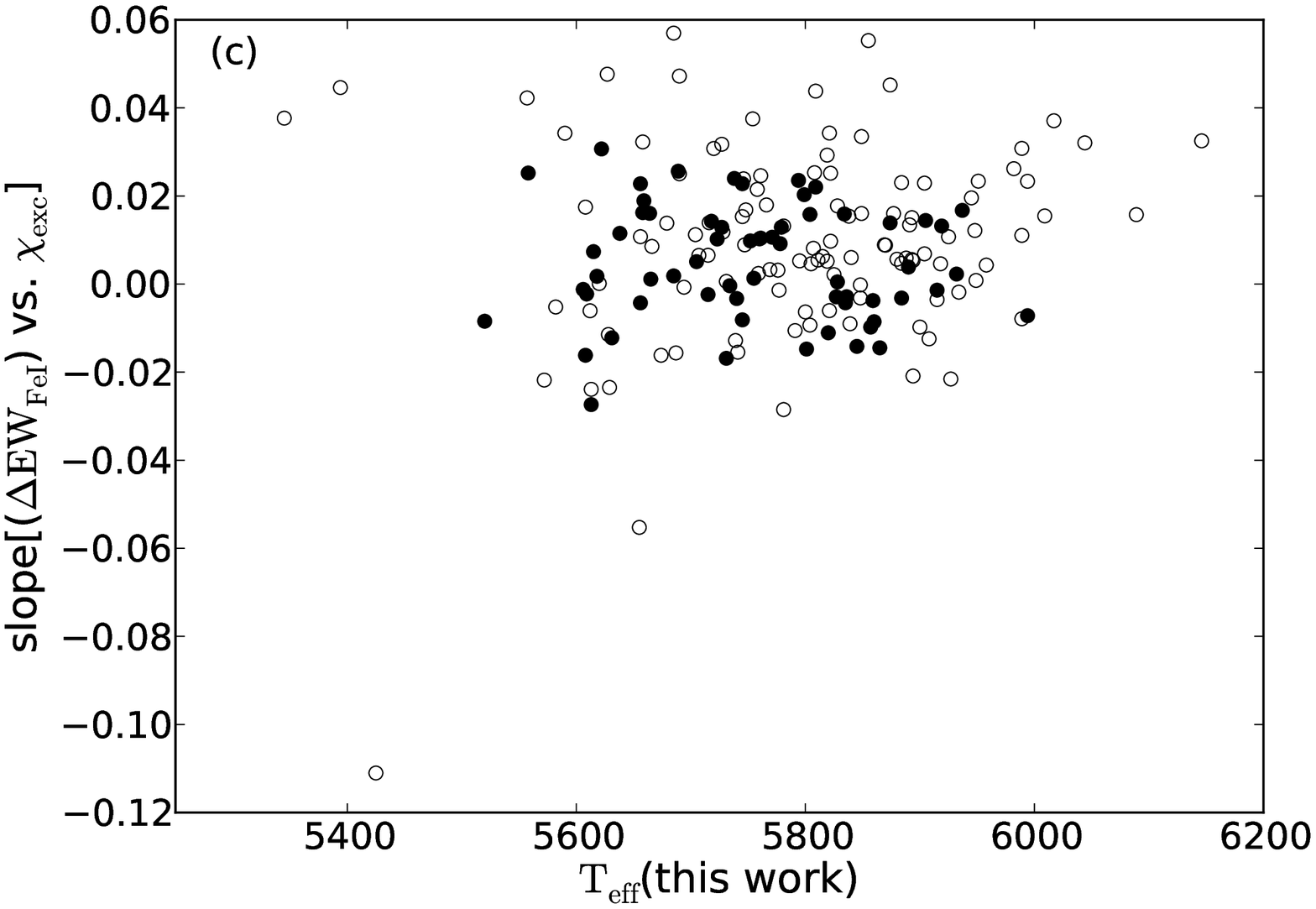}
\label{fig:subfig7}
}
\subfigure{
  \includegraphics[scale=0.35]{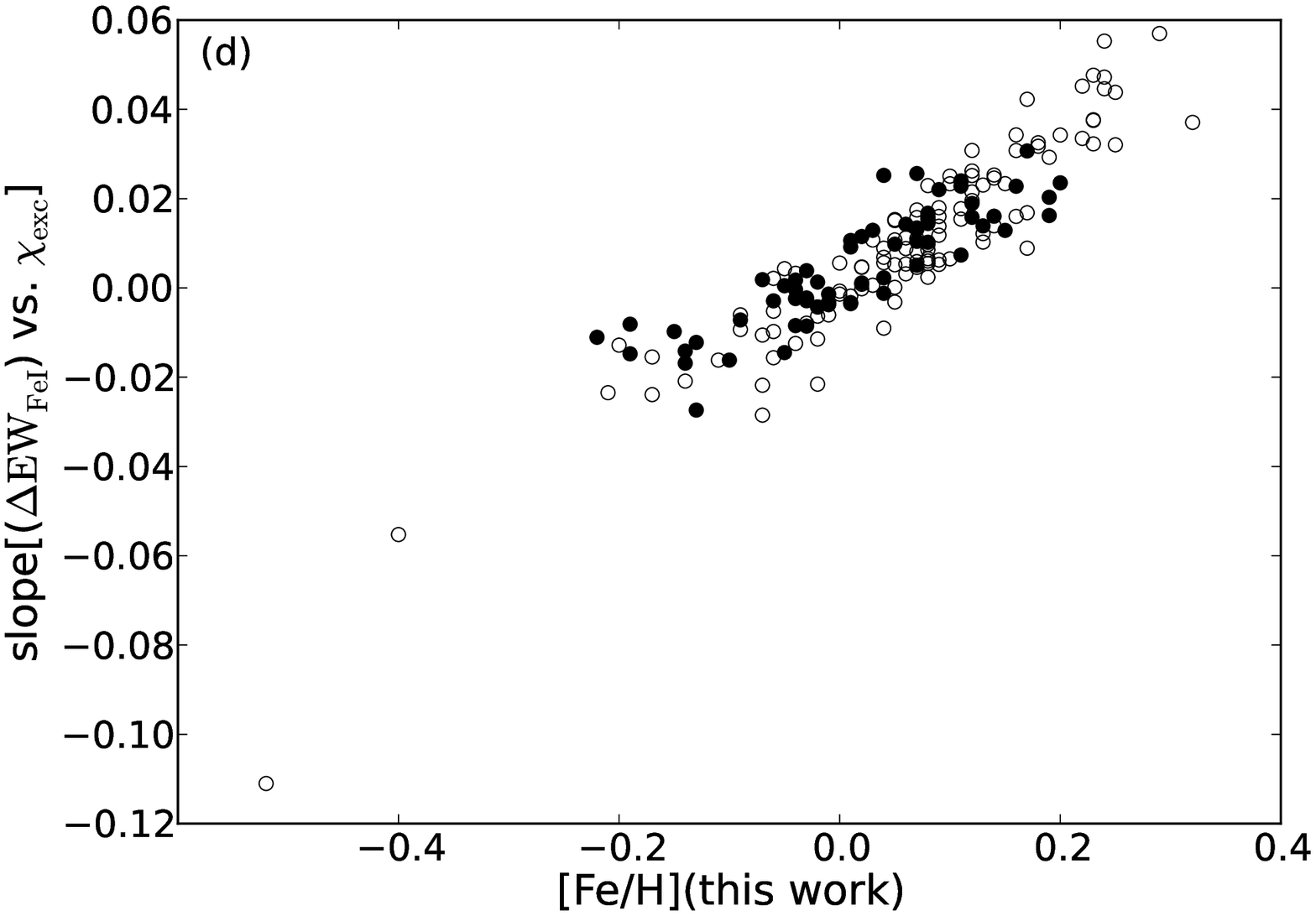}
\label{fig:subfig8}
}
\caption[Optional caption for list of figures]{Panels (a) to (d) show the trends in data we used for the degeneracy line method. In all panels the open circles show the FE12 and filled circles show the FE14 sample. Panels (a) and (b) show how the median difference in equivalent width for the iron lines $\left<\Delta \mathrm{EW}_{\mathrm{Fe\thinspace I}}\right>$ depends on effective temperature and metallicity. Panels (c) and (d) show how the slope[($\Delta\mathrm{EW}_{\mathrm{Fe\thinspace I}}$) vs. $\chi_{\mathrm{exc}}$] depends on the same parameters. Contrary to expectations, the slope parameter only depends on metallicity and not on effective temperature (as already noted in D12).}
\label{trends}
\end{figure*}

We then used the whole line list and our dedicated {\sc twospec} code (for details see D12) to determine the relevant quantities $\left<\Delta \mathrm{EW}_{\mathrm{all}}\right>$, $\left<\Delta \mathrm{EW}_{\mathrm{Fe\thinspace I}}\right>$ and slope[($\Delta\mathrm{EW}_{\mathrm{Fe\thinspace I}}$) vs. $\chi_{\mathrm{exc}}$], needed to apply our degeneracy lines methods (i) and (ii) from D12  to double-check the calibration of our new spectroscopic values. Again, our Ceres spectra were used as the solar comparison spectra for our purely differential analysis of the full FEplus sample of solar analogue stars.

The difference in relative equivalent width for a single stellar line, as compared to the solar values, is given by

\begin{equation}
  \Delta \mathrm{EW} = (\mathrm{EW}(\star)-\mathrm{EW}(\astrosun))/\mathrm{EW}(\astrosun),
\end{equation}

\noindent
where the solar values are calculated from a comparison solar spectrum, taken from the asteroid Ceres. From this the previously mentioned quantities follow as
\begin{itemize}
\item median difference in equivalent width $\left<\Delta \mathrm{EW}\right>$ of all spectral lines
        \begin{equation}
        \left<\Delta \mathrm{EW}_{\mathrm{all}}\right> = \left<(\mathrm{EW}_{\star,\mathrm{all}}-\mathrm{EW}_{\astrosun,\mathrm{all}})/\mathrm{EW}_{\astrosun,\mathrm{all}}\right>,
        \label{eq1}
        \end{equation}

\item median difference in equivalent width $\left<\Delta \mathrm{EW}\right>$ of all FeI lines
        \begin{equation}
        \left<\Delta \mathrm{EW}_{\mathrm{FeI}}\right> = \left<(\mathrm{EW}_{\star,\mathrm{FeI}}-\mathrm{EW}_{\astrosun,\mathrm{FeI}})/\mathrm{EW}_{\astrosun,\mathrm{FeI}}\right>,
        \label{eq2}
        \end{equation}

\item the slope of a linear fit to the difference in equivalent width versus the excitation potential of the line,\linebreak slope[($\Delta\mathrm{EW}_{\mathrm{Fe\thinspace I}}$) vs. $\chi_{\mathrm{exc}}$].
\end{itemize}

\noindent
Figure \ref{trends} shows the dependencies of $\left<\Delta \mathrm{EW}_{\mathrm{FeI}}\right>$ and the slope[($\Delta \mathrm{EW}_{\mathrm{FeI}}$) vs. $\chi_{\mathrm{exc}}$] on metallicity and effective temperature for the stars in the FE12 sample (open circles) and the FE14 sample (filled circles), which allows us to apply 2D least-square fitting to derive equations of the type $a \, [\mathrm{Fe/H}] + b \frac{T_{\mathrm{eff}}-5777}{5777} + c$ for the FEplus sample:

\begin{equation}
  \left<\Delta \mathrm{EW}_{\mathrm{all}}\right> =
  0.905[\mathrm{Fe/H}]-3.893\frac{T_{\mathrm{eff}}-5777}{5777}-0.016
\end{equation}

\begin{equation}
  \left<\Delta \mathrm{EW}_{\mathrm{FeI}}\right> =
  0.768[\mathrm{Fe/H}]-3.830\frac{T_{\mathrm{eff}}-5777}{5777}-0.015
\end{equation}

\begin{equation}
  \mathrm{slope}[(\Delta \mathrm{EW}_{\mathrm{FeI}})\ \mathrm{vs.}\ \chi_{\mathrm{exc}}]=
  0.179[\mathrm{Fe/H}]+0.012\frac{T_{\mathrm{eff}}-5777}{5777}
.\end{equation}

\noindent
For the Sun, the left-hand side of these equations should be zero. Therefore they can be rewritten as $[\mathrm{Fe/H}] = d \frac{T_{\mathrm{eff}}-5777}{5777} + e $

\begin{equation}
  [\mathrm{Fe/H}]_{\left<\Delta \mathrm{EW}_{\mathrm{all}}\right>} =
  4.301\frac{T_{\mathrm{eff}}-5777}{5777}+0.018
\end{equation}

\begin{equation}
  [\mathrm{Fe/H}]_{\left<\Delta \mathrm{EW}_{\mathrm{FeI}}\right>} =
  4.990\frac{T_{\mathrm{eff}}-5777}{5777}+0.020
\end{equation}

\begin{equation}
  [\mathrm{Fe/H}]_{\mathrm{slope}[(\Delta \mathrm{EW}_{\mathrm{FeI}})\
    \mathrm{vs.}\ \chi_{\mathrm{exc}}]} = -0.070\frac{T_{\mathrm{eff}}-5777}{5777}
.\end{equation}

\begin{figure}
  \centering 
  \includegraphics[scale=0.4]{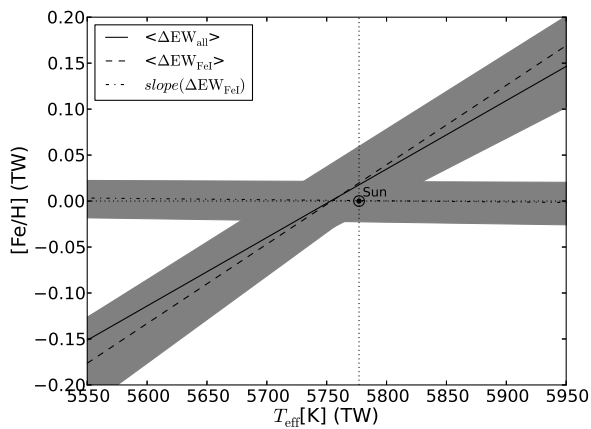}
  \caption{Solar zero point for the degeneracy lines analysis of the FEplus sample using our spectroscopic values for $T_{\mathrm{eff}}$ and [Fe/H] from this work (TW). It is clear that the crossing point of the lines falls close to the solar zero--point, thus showing that our values are well centred on the Sun. Shaded areas show the $1\sigma$ scatter in the lines.}
  \label{zero}
\end{figure}

\noindent
Figure ~\ref{zero} shows how the three different degeneracy lines suggest that the solar zero point of our spectroscopic values lies around $5755\pm40$~K in effective temperature and $0.00\pm0.02$~dex in metallicity, thus implying that our values are very well calibrated (see also Table~\ref{parameters} for the equation parameters). This shows that for our sample of solar analogue stars, spectroscopic stellar parameters are favoured over the GCS-III photometric ones, which we have already shown to have an offset in D12 and D14. 

\subsection{Comparison to literature}

To compare our results to previous work, we searched the literature for papers that give spectroscopic metallicities and effective temperatures for our sample stars. We found 16 papers with values for 76 of our stars. The papers are \citet{Fo76} (1 star), \citet{Ra99} (1 star), \citet{So06} (2 stars), \citet{Je08} (4 stars), \citet{Sb08} (12 stars), \citet{So08} (21 stars), \citet{Bu10} (2 stars), \citet{Gh10} (19 stars), \citet{Ka11} (1 star), \citet{Le11} (10 stars), \citet{Pr11} (7 stars), \citet{So11} (7 stars), \citet{Ko12} (1 star), \citet{Ma12} (9 stars), \citet{Pa12} (2 stars) and \citet{Ra14b} (21 stars).

\begin{figure*}
\centering
\subfigure{
  \includegraphics[scale=0.4]{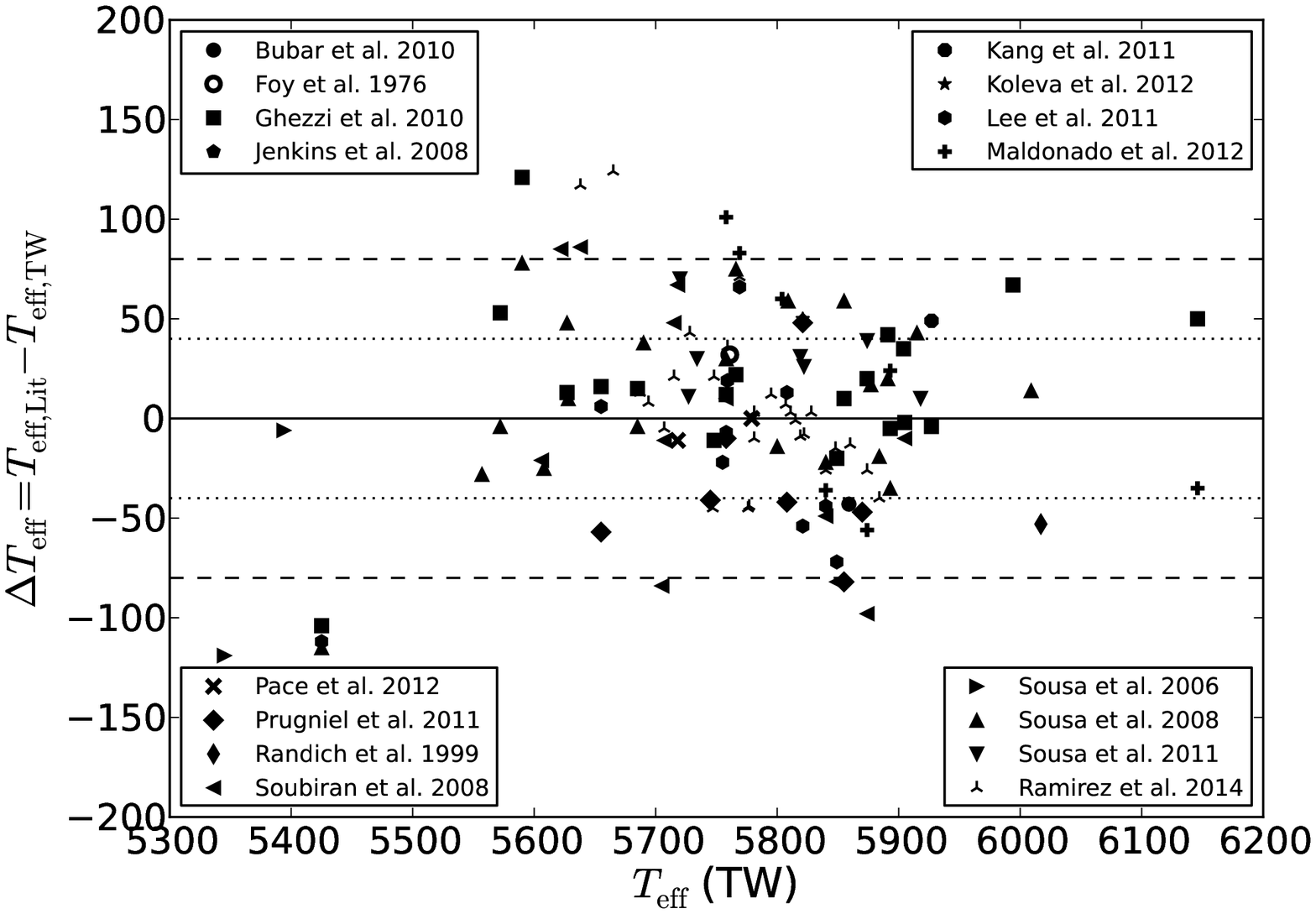}
\label{compT}
}
\subfigure{
  \includegraphics[scale=0.4]{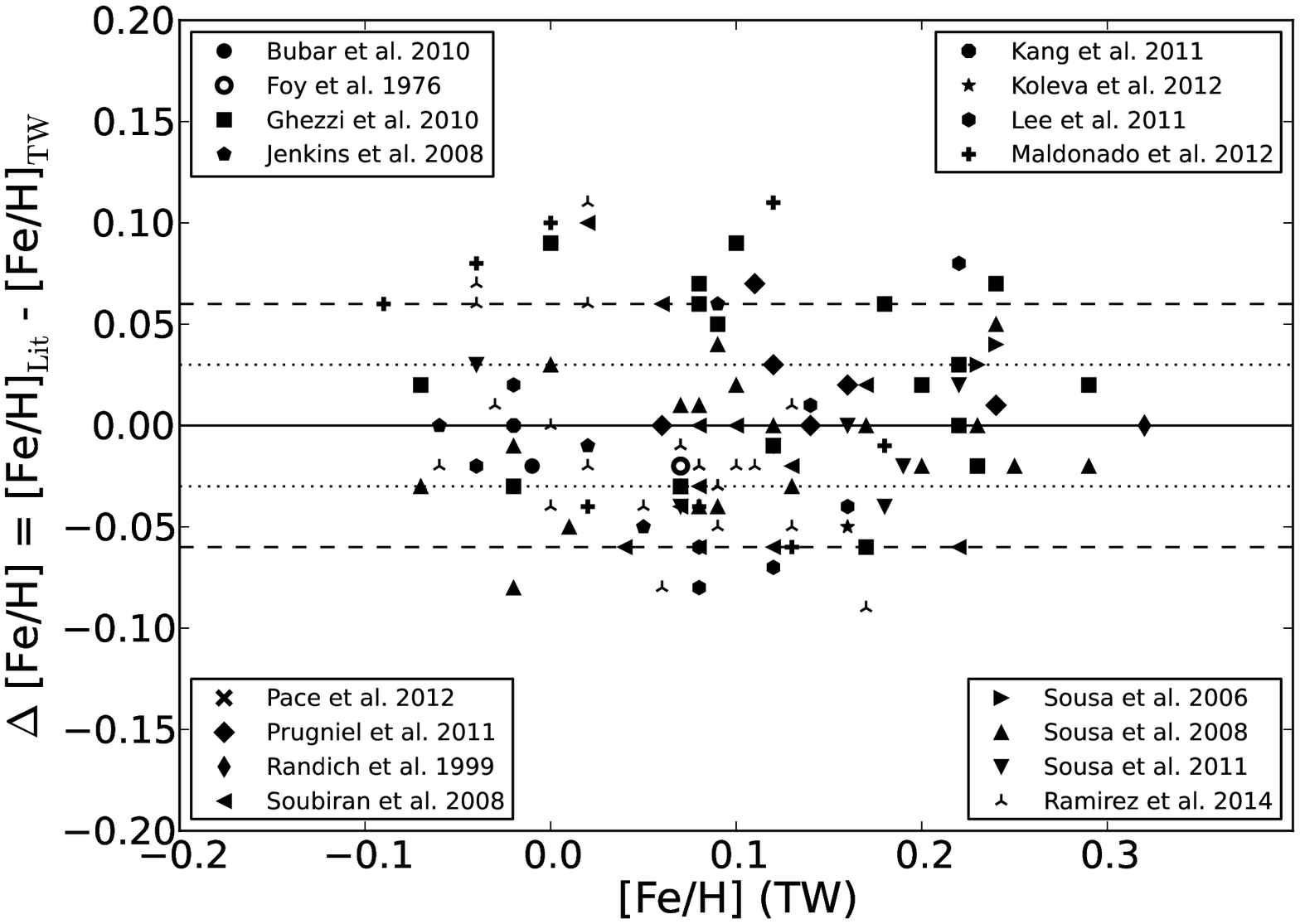}
\label{compM}
}
\caption[Optional caption for list of figures]{Figures \ref{compT} and \ref{compM} show how the spectroscopic literature values of 76 of our stars compare to our determined values for effective temperature and metallicity. Both cases show literature values (Lit) minus the values determined in this work (TW). The horizontal lines show the $1\sigma$ and $2\sigma$ scatter in our values. We find no significant trends or offsets.}
\label{littrends}
\end{figure*}

In Fig.\ref{compT} and \ref{compM} we plot the difference in effective temperature and metallicity between these literature values and our own spectroscopic values from this work. The dashed horizontal lines show the $1\sigma$ and $2\sigma$ scatter ranges for our determinations. The average difference and scatter in effective temperature is $4\pm47$~K and $0.00\pm0.05$~dex in metallicity. We find no significant offset or trend with effective temperature or metallicity. We also tested whether the difference in effective temperature shows any trend with metallicity and vice versa, but did not find any there either. For three papers, where we have $\sim$20 stars in common, we determined the average offsets for those stars, which is consistent with the overall average: For the 21 stars in common with \citet{So08} we find $\Delta T = 5\pm40$~K and $\Delta$[Fe/H] = $-0.01\pm0.03$~dex; for the 19 stars in common with \citet{Gh10} we find $\Delta T = 15\pm39$~K and $\Delta$[Fe/H] = $0.03\pm0.04$~dex; and for the 21 stars in common with \citet{Ra14b} we find $\Delta T = 7\pm42$~K and $\Delta$[Fe/H] = $-0.01\pm0.05$~dex. This shows that our values agree excellently well with the literature, and our method for determining the stellar parameters seems to be consistent.

\section{Solar twins}
\label{solartwins}

In addition to determining the new stellar parameters, we also used the new data (FE14 sample) to search for more solar twins and then to see how solar-like these stars are in terms of their currently accepted parameters. Considering that there is a variety of solar twin definitions in the literature, we specify that in this paper, as in D12 (and D14), we consider to be solar twins all stars that are indistinguishable from the Sun within the errors, with respect to the inspected spectroscopic features. In our case, these are (differences in) EW of selected lines that ultimately trace effective temperature and metallicity.

\subsection{Methods and results}

We used method (i) and method (ii) and the same line list for the FE14 sample as we did for the FE12 sample in D12 for consistency with our previous study. We used our {\sc twospec} code to determine the differences in equivalent width for all 109 spectral lines in this line list \citep[Ra09, ][]{Ra09}, which has lines from 20 different elements (O\thinspace I, Na\thinspace I, Mg\thinspace I, Al\thinspace I, Si\thinspace I, S\thinspace I, K\thinspace I, Ca\thinspace I, Sc\thinspace II, Ti\thinspace I, V\thinspace I, Cr\thinspace I, Mn\thinspace I, Fe\thinspace I+II, Co\thinspace I, Ni\thinspace I, Cu\thinspace I, Zn\thinspace I, Zr\thinspace II and Ba\thinspace II) over a wavelength range of $5000 - 8000\AA\ $. 
\\
\\
In method (i) we calculated the median $\left<\Delta \mathrm{EW}_{\mathrm{all}}\right>$ (Eq.~\ref{eq1}) and scatter $\chi^{2}(\Delta \mathrm{EW}_{\mathrm{all}})$ of the differences in EWs for all 109 lines in the list, which is closely related to the first criterion in \citet{Me06}. A solar twin should be indistinguishable from the Sun and accordingly have no differences in EW, resulting in a median difference in EW of zero and a $\chi^{2}$ also of zero. We define a twin to be a star in method (i) if the median difference in EW for all lines vanishes within the observational error and the scatter is minimal:  $\left<\Delta \mathrm{EW}_{\mathrm{all}}\right>$ = 0 within 2$\sigma$ and $\chi^{2}(\Delta \mathrm{EW}_{\mathrm{all}}) \leq 2$ (the $\chi^{2}$ limit for the FE14 sample is higher than for the FE12 sample, as the reduced and extracted spectra from the former showed more noise due to the inconsistencies in the reduction process).

This gave us four solar twins: HD25680, HD39649, HD41708, and HD76440.
\\
\\
In method (ii), which is based on \citet{Me07}, we concentrated on using only the 33 Fe\thinspace I lines in the list and determined the median difference in EW (Eq.~\ref{eq2}), and the fit of the difference in EW of every Fe\thinspace I line versus its excitation potential ($\chi_{\mathrm{exc}}$) and the corresponding slope. For a solar twin, both of these quantities should vanish, and in our analysis we call a solar twin all stars that have $\left<\Delta \mathrm{EW}_{\mathrm{FeI}}\right>$ and a slope[($\Delta \mathrm{EW}_{\mathrm{FeI}}$) vs. $\chi_{\mathrm{exc}}$] of zero within $2\sigma$.

This method gave us five twins: HD29601, HD35769, HD39649, HD67010, and HD76440.
\\
\\
In total, this resulted in seven solar twins in the FE14 sample, one of which (HD76440) we already identified in D14. Of the other four twins, three twins are new and previously unpublished. They are HD29601, HD39649, and HD67010. 

In Table \ref{twins} we list our full set of twins (from the FE12 and FE14 samples), showing their GCS-III parameters and the parameters determined in this work, as well as our determined values for the surface gravity.

\begin{table*}
\caption{List of solar twins in both samples. Shown are the GCS-III values and the derived values from this work (TW). We also add our values for log g, as it is also well centred on the solar value of 4.44. Two stars in italics are twins that have only
been identified by using other line lists than the Ra09 list adopted in D12, see Sect.~\ref{linetwin}. Stars with a ($^{*}$) are considered the best twins in this analysis, see Sect.~\ref{asteroids} and Fig.~\ref{asttwins}.}
\label{twins}
\centering
\begin{tabular}{lcccccc}
\hline
& HD number & $T_{\mathrm{eff}}$ (GCS) & $T_{\mathrm{eff}}$ (TW) & [Fe/H] (GCS) & [Fe/H] (TW) & log g (TW)\\
& & [K] & [K] & [dex] & [dex] & [dex]\\
\hline
FE12 sample & $78660^{*}$ & 5715 & 5776 & $-$0.09 & \hphantom{$-$}0.06 & 4.55\\
& $97356^{*}$ & 5754 & 5805 & $-$0.06 & \hphantom{$-$}0.02 & 4.35\\
& 117860 & 5821 & 5948 & $-$0.08 & \hphantom{$-$}0.13 & 4.62\\
& 126525 & 5585 & 5628 & $-$0.19 & $-$0.02 & 4.52\\
& $138573^{*}$ & 5689 & 5777 & $-$0.10 & \hphantom{$-$}0.00 & 4.46\\
& 142415 & 5916 & 5904 & \hphantom{$-$}0.04 & \hphantom{$-$}0.08 & 4.40\\
& $146233^{*}$ & 5768 & 5819 & $-$0.02 & \hphantom{$-$}0.08 & 4.47\\
& 147513 & 5781 & 5885 & $-$0.13 & \hphantom{$-$}0.04 & 4.53\\
& $163441^{*}$ & 5702 & 5795 & $-$0.09 & \hphantom{$-$}0.09 & 4.43\\
& 173071 & 5875 & 6044 & $-$0.04 & \hphantom{$-$}0.25 & 4.49\\
\hline
FE14 sample & 25680 & 5781 & 5905 & $-$0.08 & \hphantom{$-$}0.08 & 4.57\\
& ${\it 28068^{*}}$ & {\it 5728} & {\it 5761} & {\it $-$0.04} & {\it \hphantom{$-$}0.07} & {\it 4.32}\\
& 29601 & 5598 & 5608 & $-$0.14 & $-$0.10 & 4.24\\
& 35769 & 5649 & 5631 & $-$0.12 & $-$0.13 & 4.34\\
& $39649^{*}$ & 5675 & 5740 & $-$0.09 & \hphantom{$-$}0.01 & 4.40\\
& 41708 & 5821 & 5928 & $-$0.07 & \hphantom{$-$}0.08 & 4.45\\
& 67010 & 5598 & 5613 & $-$0.20 & $-$0.13 & 4.48\\
& $76440^{*}$ & 5623 & 5781 & $-$0.23 & $-$0.05 & 4.34\\
& ${\it 77461^{*}}$ & {\it 5754} & {\it 5835} & {\it $-$0.18} & {\it $-$0.02} & {\it 4.49}\\
\hline
average $\pm$ scatter & & $5727\pm100$ & $5799\pm128$ & $-0.10\pm0.07$ & \hphantom{$-$}$0.03\pm0.10$ & $4.45\pm0.10$\\
\hline
\end{tabular}
\end{table*}

Our spectroscopic values are hotter and more metal rich and better centred on the Sun, with an average and scatter of the twin parameters of $5799\pm128$~K and $0.03\pm0.10$~dex, as opposed to the average and scatter of the GCS-III values of $5727\pm100$~K and $-0.10\pm0.07$~dex. This conclusion on the respective zero--point is only based on 17 stars; a more reliable test is based on the full FEplus sample of solar analogue stars, which can be found in the next section.

\subsection{Line list dependance on solar twin determination}
\label{linetwin}

One test of consistency was to use different line lists for our twin search in the FE14 sample, as it is currently appreciated that this choice might influence spectroscopic results \citep[e.g.][]{So14, Jo14}. We chose four different lists, some using mainly iron lines, others using lines of many different elements and species. They were reported by \citet{Ra09} (Ra09), \citet{Bi12} (Bi12), \citet{Be03} (Be03), and \citet{So08} (So08). In addition, we also compiled a combined list (comb) of all four. This resulted in five different line lists of different lengths, ranging from 94 to 447 entries. (For details on the lists see Tables \ref{details}, \ref{overlap} and \ref{linelists}).

\begin{table}
\caption{Line list details, amount of lines per element and species}
\label{details}
\centering
\begin{tabular}{lccccc}
\hline
& Ra09 & Bi12 & Be03 & So08 & comb\\
\hline
O\thinspace I & 3 & $-$ & $-$ & $-$ & 3\\
Na\thinspace I & 2 & 3 & 3 & $-$ & 4\\
Mg\thinspace I & 3 & $-$ & 2 & $-$ & 4\\
Al\thinspace I & 4 & 1 & $-$ & $-$ & 4\\
Si\thinspace I & 8 & 7 & 15 & $-$ & 19\\
Si\thinspace II & $-$ & $-$ & 13 & $-$ & 13\\
S\thinspace I & 2 & $-$ & $-$ & $-$ & 2\\ 
K\thinspace I & 1 & $-$ & $-$ & $-$ & 1\\ 
Ca\thinspace I & 8 & 13 & 20 & $-$ & 23\\
Ca\thinspace II & $-$ & $-$ & 1 & $-$ & 1\\
Ti\thinspace I & 6 & 9 & 15 & $-$ & 22\\
Ti\thinspace II & $-$ & $-$ & 8 & $-$ & 8\\
Cr\thinspace I & 5 & 7 & 6 & $-$ & 14\\
Cr\thinspace II & $-$ & $-$ & 3 & $-$ & 3\\ 
Fe\thinspace I & 33 & 80 & 105 & 192 & 248\\
Fe\thinspace II & 7 & 7 & 15 & 25 & 28\\
Co\thinspace I & 2 & $-$ & $-$ & $-$ & 2\\
Ni\thinspace I & 9 & 22 & 29 & $-$ & 44\\
Ni\thinspace II & $-$ & $-$ & 1 & $-$ & 1\\
Cu\thinspace I & 1 & $-$ & $-$ & $-$ & 1\\
Zn\thinspace I & $-$ & 1 & 2 & $-$ & 2\\
\hline
total & 94 & 151 & 238 & 217 & 447\\
\hline
\end{tabular}
\end{table}

\begin{table}
\caption{Number of common lines in the different line lists.}
\label{overlap}
\centering
\begin{tabular}{cccc}
\hline
& Bi12 & Be03 & So08\\
\hline
Ra09 & 30 & 56 & 36\\
Bi12 & $-$ & 60 & 57\\
Be03 & $-$ & $-$ & 90\\
\hline
\end{tabular}
\end{table}

\begin{table}
\caption{Example of spectral lines used in the four line lists, sorted by element. Full table available at the CDS.}
\label{linelists}
\centering
\begin{tabular}{lccccc}
\hline
Element & wavelength [\AA] & Ra09 & Bi12 & Be03 & So08\\
\hline
O\thinspace I & 7771.95 & x & & &\\                     
O\thinspace I & 7774.17 & x & & &\\     
O\thinspace I & 7775.39 & x & & &\\
Na\thinspace I & 5682.63 & & x & x & \\
Na\thinspace I & 5688.22 & & & x & \\
Na\thinspace I & 6154.23 & x & x & & \\
Na\thinspace I & 6160.75 & x & x & x & \\
Mg\thinspace I & 4571.10 & & & x & \\
Mg\thinspace I & 5711.09 & x & & x & \\
Mg\thinspace I & 6318.71 & x & & & \\
\vdots & \vdots & \vdots & \vdots & \vdots & \vdots \\
\hline
\end{tabular}
\end{table}

In Table~\ref{twinslists} we list which twins we found using the different line lists from the FE14 sample. We show that there are some differences in the results, with five of the nine twins being recovered by most of the line lists, but four twins only present in one or two line lists. This shows that there is still a certain degree of arbitrariness in choosing what to consider a solar twin, depending on the line list used, for
instance. We call all of them solar twins here, meaning that
we do not favour any of the used line lists. There are two twins in Table~\ref{twinslists} that are marked in italics in Table~\ref{twins}
because they were found by using other line lists than Ra09 and are thus not fully consistent with the methods used before in D12. The twins marked with a ($^{*}$) are considered our best twins in this work and are discussed in Sect.~\ref{asteroids}.

\begin{table}
\caption{Solar twins determined using different line lists.}
\label{twinslists}
\centering
\begin{tabular}{lcccccc}
\hline
HD & Ra09 & Bi12 & Be03 & So08 & comb & total\\
\hline
25680 & yes & yes & yes & no & yes & 4\\
28068 & no & yes & yes & yes & yes & 4\\
29601 & yes & no & no & no & yes & 2\\
35769 & yes & yes & yes & yes & yes & 5\\
39649 & yes & no & no & no & no & 1\\
41708 & yes & yes & yes & no & yes & 4\\
67010 & yes & yes & no & no & no & 2\\
76440 & yes & yes & yes & yes & yes & 5\\
77461 & no & yes & no & no & no & 1\\
\hline
\end{tabular}
\end{table}

\section{GCS tests in the FE14 and FEplus samples}

In D12 we introduced our degeneracy line method to test the solar zero--point of the GCS catalogue and found offsets of $-97\pm35$~K and $-0.12\pm0.02$~dex in effective temperature and metallicity, respectively, similar to those we found from solar twin arguments in the previous section. D12 used the FE12 sample; here we repeat the analysis with the FE14 and the FEplus samples. 

\begin{table*}
\caption{List of parameters $a$ to $e$ for $a \, [\mathrm{Fe/H}] + b \frac{T_{\mathrm{eff}}-5777}{5777} + c$ and $[\mathrm{Fe/H}] = d \frac{T_{\mathrm{eff}}-5777}{5777} + e$ as in Eqs.~$4-9$. The parameter source can be either this work (TW), GCS-III, or C11.}
\label{parameters}
\centering
\begin{tabular}{lllcccccc}
\hline
Sample & Line & Param. & $a$ & $b$ & $c$ & $d$ & $e$ & Figure\\
& list & source &&&&&&\\
\hline
FEplus & Bi12 & TW & \hphantom{$-$}$0.905\pm0.136$ & $-3.893\pm1.168$ & $-0.016\pm0.013$ & \hphantom{$-$}$4.301\pm1.462$ & \hphantom{$-$}$0.018\pm0.015$ & $3$\\
FEplus & Bi12 & TW & \hphantom{$-$}$0.768\pm0.115$ & $-3.830\pm1.149$ & $-0.015\pm0.012$ & \hphantom{$-$}$4.990\pm1.670$ & \hphantom{$-$}$0.020\pm0.016$ & $3$\\
FEplus & Bi12 & TW & \hphantom{$-$}$0.179\pm0.027$ & \hphantom{$-$}$0.012\pm0.004$ & \hphantom{$-$}$0.000\pm0.010$ & $-0.070\pm0.024$ & \hphantom{$-$}$0.001\pm0.001$ & $3$\\
&&&&&&&\\
FE12 & Ra09 & GCS & \hphantom{$-$}$1.056\pm0.158$ & $-3.829\pm1.149$ & \hphantom{$-$}$0.066\pm0.053$ & \hphantom{$-$}$3.625\pm1.235$ & $-0.062\pm0.050$ & $5$\\
FE12 & Ra09 & GCS & \hphantom{$-$}$0.856\pm0.128$ & $-3.769\pm1.131$ & \hphantom{$-$}$0.045\pm0.036$ & \hphantom{$-$}$4.406\pm1.498$ & $-0.053\pm0.043$ & $5$\\
FE12 & Ra09 & GCS & \hphantom{$-$}$0.256\pm0.038$ &  & \hphantom{$-$}$0.027\pm0.022$ & $ $ & $-0.106\pm0.015$ & $5$\\
&&&&&&&\\
FE14 & Ra09 & GCS & \hphantom{$-$}$1.384\pm0.208$ & $-4.035\pm1.211$ & \hphantom{$-$}$0.055\pm0.044$ & \hphantom{$-$}$2.916\pm0.991$ & $-0.040\pm0.032$ & $5~\&~6$\\
FE14 & Ra09 & GCS & \hphantom{$-$}$1.372\pm0.206$ & $-4.817\pm1.445$ & \hphantom{$-$}$0.042\pm0.034$ & \hphantom{$-$}$3.510\pm1.193$ & $-0.031\pm0.025$ & $5~\&~6$\\
FE14 & Ra09 & GCS & \hphantom{$-$}$0.173\pm0.026$ & \hphantom{$-$}$0.221\pm0.066$ & \hphantom{$-$}$0.017\pm0.014$ & $-1.279\pm0.435$ & $-0.096\pm0.078$ & $5~\&~6$\\
&&&&&&&\\
FEplus & Ra09 & GCS & \hphantom{$-$}$1.206\pm0.181$ & $-4.052\pm1.2156$ & \hphantom{$-$}$0.058\pm0.046$ & \hphantom{$-$}$3.360\pm1.142$ & $-0.048\pm0.039$ & $5$\\ 
FEplus & Ra09 & GCS & \hphantom{$-$}$1.076\pm0.161$ & $-4.325\pm1.298$ & \hphantom{$-$}$0.040\pm0.032$ & \hphantom{$-$}$4.018\pm1.366$ & $-0.037\pm0.030$ & $5$\\ 
FEplus & Ra09 & GCS & \hphantom{$-$}$0.231\pm0.035$ & \hphantom{$-$}$0.158\pm0.047$ & \hphantom{$-$}$0.024\pm0.019$ & $-0.682\pm0.232$ & $-0.103\pm0.083$ & $5$\\ 
&&&&&&&\\
FE14 & Bi12 & GCS & \hphantom{$-$}$1.418\pm0.213$ & $-3.817\pm1.145$ & \hphantom{$-$}$0.072\pm0.058$ & \hphantom{$-$}$2.691\pm0.915$ & $-0.051\pm0.041$ & $6$\\
FE14 & Bi12 & GCS & \hphantom{$-$}$1.417\pm0.197$ & $-3.808\pm1.142$ & \hphantom{$-$}$0.072\pm0.058$ & \hphantom{$-$}$2.687\pm0.914$ & $-0.051\pm0.039$ & $6$\\
FE14 & Bi12 & GCS & \hphantom{$-$}$0.169\pm0.025$ & \hphantom{$-$}$1.161\pm0.348$ & \hphantom{$-$}$0.036\pm0.029$ & $-6.892\pm2.343$ & $-0.214\pm0.173$ & $6$\\
&&&&&&&\\
FE14 & Be03 & GCS & \hphantom{$-$}$1.468\pm0.220$ & $-3.511\pm1.053$ & \hphantom{$-$}$0.071\pm0.057$ & \hphantom{$-$}$2.392\pm0.813$ & $-0.048\pm0.039$ & $6$\\
FE14 & Be03 & GCS & \hphantom{$-$}$1.442\pm0.216$ & $-4.555\pm1.367$ & \hphantom{$-$}$0.053\pm0.042$ & \hphantom{$-$}$3.158\pm1.074$ & $-0.037\pm0.030$ & $6$\\
FE14 & Be03 & GCS & \hphantom{$-$}$0.154\pm0.023$ & \hphantom{$-$}$0.204\pm0.061$ & \hphantom{$-$}$0.017\pm0.014$ & $-1.323\pm0.450$ & $-0.112\pm0.091$ & $6$\\
&&&&&&&\\
FE14 & So08 & GCS & \hphantom{$-$}$1.451\pm0.218$ & $-2.963\pm0.889$ & \hphantom{$-$}$0.076\pm0.061$ & \hphantom{$-$}$2.043\pm0.695$ & $-0.053\pm0.043$ & $6$\\
FE14 & So08 & GCS & \hphantom{$-$}$1.425\pm0.214$ & $-3.634\pm1.090$ & \hphantom{$-$}$0.066\pm0.053$ & \hphantom{$-$}$2.550\pm0.867$ & $-0.046\pm0.037$ & $6$\\
FE14 & So08 & GCS & \hphantom{$-$}$0.151\pm0.023$ & \hphantom{$-$}$0.161\pm0.048$ & \hphantom{$-$}$0.017\pm0.014$ & $-1.064\pm0.362$ & $-0.112\pm0.091$ & $6$\\
&&&&&&&\\
FE14 & comb & GCS & \hphantom{$-$}$1.434\pm0.215$ & $-3.320\pm0.996$ & \hphantom{$-$}$0.071\pm0.057$ & \hphantom{$-$}$2.316\pm0.787$ & $-0.049\pm0.032$ & $6$\\
FE14 & comb & GCS & \hphantom{$-$}$1.448\pm0.223$ & $-3.829\pm1.149$ & \hphantom{$-$}$0.064\pm0.051$ & \hphantom{$-$}$2.645\pm0.899$ & $-0.044\pm0.036$ & $6$\\
FE14 & comb & GCS & \hphantom{$-$}$0.168\pm0.025$ & \hphantom{$-$}$0.116\pm0.035$ & \hphantom{$-$}$0.019\pm0.015$ & $-0.695\pm0.236$ & $-0.113\pm0.092$ & $6$\\
&&&&&&&\\
FE14 & Ra09 & TW & \hphantom{$-$}$0.694\pm0.097$ & $-3.167\pm0.950$ & $-0.010\pm0.008$ & \hphantom{$-$}$4.562\pm0.991$ & \hphantom{$-$}$0.015\pm0.012$ & $7$\\
FE14 & Ra09 & TW & \hphantom{$-$}$0.683\pm0.096$ & $-3.204\pm1.057$ & $-0.014\pm0.011$ & \hphantom{$-$}$4.691\pm1.548$ & \hphantom{$-$}$0.020\pm0.016$ & $7$\\
FE14 & Ra09 & TW & \hphantom{$-$}$0.164\pm0.025$ & \hphantom{$-$}$0.079\pm0.026$ & \hphantom{$-$}$0.000\pm0.011$ & $-0.485\pm0.160$ & $-0.001\pm0.011$ & $7$\\
&&&&&&&\\
FE14 & Bi12 & TW & \hphantom{$-$}$0.672\pm0.101$ & $-2.747\pm0.907$ & \hphantom{$-$}$0.002\pm0.008$ & \hphantom{$-$}$4.087\pm1.349$ & $-0.004\pm0.009$ & $7$\\
FE14 & Bi12 & TW & \hphantom{$-$}$0.683\pm0.102$ & $-2.694\pm0.889$ & \hphantom{$-$}$0.002\pm0.009$ & \hphantom{$-$}$3.945\pm1.302$ & $-0.002\pm0.008$ & $7$\\
FE14 & Bi12 & TW & \hphantom{$-$}$0.164\pm0.025$ & \hphantom{$-$}$0.687\pm0.227$ & \hphantom{$-$}$0.012\pm0.010$ & $-4.185\pm1.381$ & $-0.073\pm0.058$ & $7$\\
&&&&&&&\\
FE14 & Be03 & TW & \hphantom{$-$}$0.555\pm0.083$ & $-2.340\pm0.772$ & \hphantom{$-$}$0.001\pm0.012$ & \hphantom{$-$}$4.212\pm1.390$ & $-0.001\pm0.009$ & $7$\\
FE14 & Be03 & TW & \hphantom{$-$}$0.557\pm0.084$ & $-2.583\pm0.852$ & $-0.003\pm0.010$ & \hphantom{$-$}$4.639\pm1.531$ & \hphantom{$-$}$0.006\pm0.011$ & $7$\\
FE14 & Be03 & TW & \hphantom{$-$}$0.144\pm0.022$ & \hphantom{$-$}$0.081\pm0.027$ & \hphantom{$-$}$0.003\pm0.012$ & $-0.567\pm0.187$ & $-0.020\pm0.016$ & $7$\\
&&&&&&&\\
FE14 & So08 & TW & \hphantom{$-$}$0.604\pm0.091$ & $-2.135\pm0.705$ & \hphantom{$-$}$0.001\pm0.011$ & \hphantom{$-$}$3.535\pm1.167$ & $-0.001\pm0.008$ & $7$\\
FE14 & So08 & TW & \hphantom{$-$}$0.617\pm0.093$ & $-2.397\pm0.791$ & \hphantom{$-$}$0.000\pm0.009$ & \hphantom{$-$}$3.882\pm1.281$ & \hphantom{$-$}$0.000\pm0.011$ & $7$\\
FE14 & So08 & TW & \hphantom{$-$}$0.144\pm0.022$ & \hphantom{$-$}$0.032\pm0.011$ & \hphantom{$-$}$0.003\pm0.011$ & $-0.221\pm0.073$ & $-0.021\pm0.017$ & $7$\\
&&&&&&&\\
FE14 & comb & TW & \hphantom{$-$}$0.594\pm0.083$ & $-2.334\pm0.770$ & \hphantom{$-$}$0.001\pm0.010$ & \hphantom{$-$}$3.930\pm1.297$ & $-0.001\pm0.009$ & $7$\\
FE14 & comb & TW & \hphantom{$-$}$0.596\pm0.089$ & $-2.413\pm0.796$ & $-0.001\pm0.011$ & \hphantom{$-$}$4.051\pm1.337$ & \hphantom{$-$}$0.002\pm0.008$ & $7$\\
FE14 & comb & TW & \hphantom{$-$}$0.157\pm0.024$ & \hphantom{$-$}$0.024\pm0.008$ & \hphantom{$-$}$0.005\pm0.005$ & $-0.151\pm0.050$ & $-0.029\pm0.023$ & $7$\\
&&&&&&&\\
CLBR & Ra09 & C11 & \hphantom{$-$}$1.263\pm0.189$ & $-3.468\pm1.040$ & $-0.004\pm0.003$ & \hphantom{$-$}$2.746\pm0.934$ & \hphantom{$-$}$0.003\pm0.002$ & $8$\\
CLBR & Ra09 & C11 & \hphantom{$-$}$1.148\pm0.172$ & $-3.949\pm1.185$ & $-0.033\pm0.026$ & \hphantom{$-$}$3.439\pm1.169$ & \hphantom{$-$}$0.028\pm0.023$ & $8$\\
CLBR & Ra09 & C11 & \hphantom{$-$}$0.182\pm0.027$ & \hphantom{$-$}$0.348\pm0.104$ & \hphantom{$-$}$0.002\pm0.001$ & $-1.912\pm0.650$ & $-0.009\pm0.007$ & $8$\\
&&&&&&&\\
IRFM & Ra09 & C11 & \hphantom{$-$}$0.985\pm0.148$ & $-3.118\pm0.935$ & \hphantom{$-$}$0.023\pm0.018$ & \hphantom{$-$}$3.163\pm1.075$ & $-0.023\pm0.019$ & $8$\\
IRFM & Ra09 & C11 & \hphantom{$-$}$0.872\pm0.131$ & $-3.113\pm0.934$ & \hphantom{$-$}$0.018\pm0.014$ & \hphantom{$-$}$3.570\pm1.214$ & $-0.021\pm0.017$ & $8$\\
IRFM & Ra09 & C11 & \hphantom{$-$}$0.211\pm0.017$ & \hphantom{$-$}$0.100\pm0.030$ & \hphantom{$-$}$0.003\pm0.002$ & $-0.477\pm0.162$ & $-0.014\pm0.011$ & $8$\\
\hline
\end{tabular}
\end{table*}

\begin{figure}
  \centering 
  \includegraphics[scale=0.4]{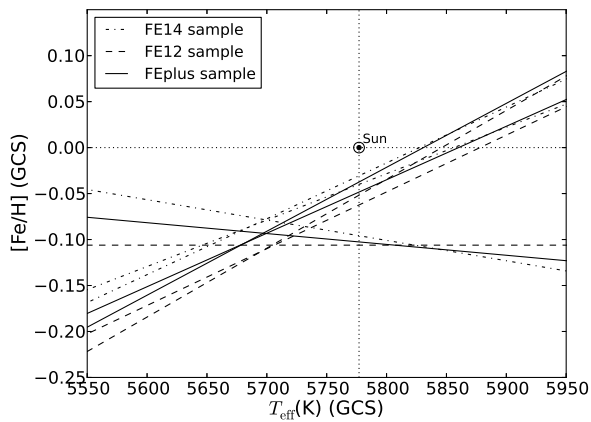}
  \caption{Degeneracy lines for the FE12, FE14, and FEplus samples, using GCS-III values for effective temperature and metallicity. Clearly, the intersection of the lines does not change with sample within the errors. For clarity, we do not plot the error bands.}
  \label{GCS}
\end{figure}

The resulting degeneracy lines can be seen in Fig.\ref{GCS} (see also Fig. 14 from D12), the corresponding equation parameters are listed in Table~\ref{parameters}; the crossing of the lines indicates the location of the solar zero--point in the GCS catalogue. It is clear from the plot that the offsets we find are independent of the specific sample used and are therefore reliable.

\subsection{Line list dependance of the degeneracy line method}

\begin{figure}
  \centering 
  \includegraphics[scale=0.4]{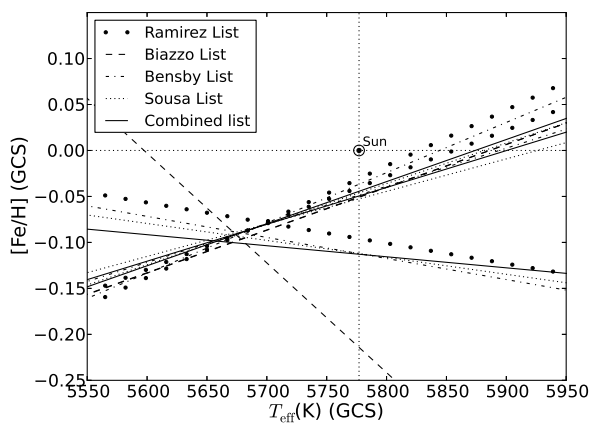}
  \caption{Crossing points for the different line lists clearly vary far less than the offset with respect to the solar reference values, thus confirming our results independently of the line list. For clarity, we do not plot the error bands.}
  \label{fig:comp}
\end{figure}

Another test of consistency was to apply our degeneracy line method to the FE14 sample stars, using all five lists separately. The resulting lines are plotted in Fig.\ref{fig:comp} (parameters in Table~\ref{parameters}). 

\begin{table}
\caption{Crossing points in Figs.~\ref{fig:comp} and~\ref{compmine} for the different lines lists.}
\label{crossing}
\centering
\begin{tabular}{llcc}
\hline
Param. & Line list & $T_{\mathrm{eff}}$ [K] & [Fe/H] [dex]\\
source & & crossing point & crossing point\\
\hline
GCS & Ra09 & 5701 & $-$0.08\\
& Bi12 & 5678 & $-$0.10\\
& Be03 & 5679 & $-$0.09\\
& So08 & 5672 & $-$0.09\\
& combined & 5657 & $-$0.10\\
\hline
average & $\pm$ scatter & $5677\pm16$ & $-0.09\pm0.01$\\
\hline
TW & Ra09 & 5755 & \hphantom{$-$}0.00\\
& Bi12 & 5728 & $-$0.04\\
& Be03 & 5751 & $-$0.02\\
& So08 & 5748 & $-$0.02\\
& combined & 5737 & $-$0.03\\
\hline
average & $\pm$ scatter & $5744\pm11$ & $-0.02\pm0.02$\\
\hline
\end{tabular}
\end{table}

It is clear that the individual intersections of lines for the different line lists fall very close to one another (see Table~\ref{crossing}). We conclude that the exact line list does not strongly influence our analysis and overall determination of the solar zero--point within the catalogue --- as long as the used lines are weak, unblended, and not saturated with a clear continuum around them.

\begin{figure}
  \centering 
  \includegraphics[scale=0.4]{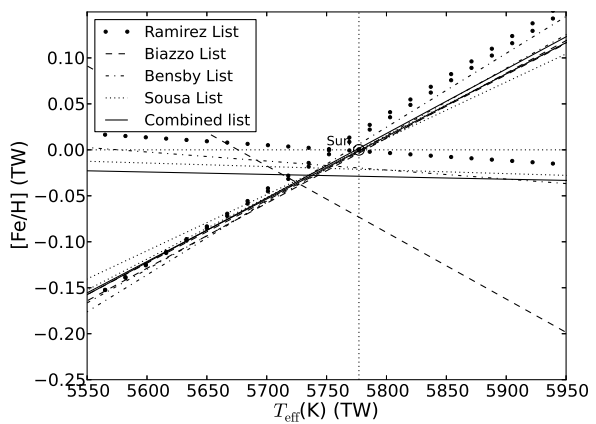}
  \caption{Degeneracy lines for the FE14 sample, using parameters determined in this work (TW) and all five line lists. For clarity, we do not plot the error bands.}
  \label{compmine}
\end{figure}

Similarly, for consistency we also applied the same analysis using the spectroscopic parameters we determined in Sect.~\ref{specparam}. Again, we found the crossing points in a narrow window of stellar parameters, giving an average effective temperature of $5744\pm40$~K and metallicity of $-0.02\pm0.02$~dex, fully consistent with the results in Sect.~\ref{deglin}. Figure~\ref{compmine} (see also Table~\ref{parameters}) shows that our analysis is independent of the line list used and that our parameters are well centred on the solar values.

\subsection{Internal consistency of the C11 reanalysis}

Recently, \citet{Ca14} tested photometric temperature scales versus the more recent interferometric data, showing that although interferometry is becoming increasingly more accurate and providing larger numbers of observable and measurable stars, it is still somewhat prone to systematics, enough to not fully settle the debate on the photometric scale.

They compared the two alternative temperature scales of the original GCS--III catalogue \citep{Ho09} and its reanalysis \citep[C11 hereafter]{Ca11} with interferometric stellar effective temperatures to determine which scale is closer to the fundamental one. This showed that both scales are within the errors of the interferometric scale ($\pm50$~K), C11 being slightly hotter, GCS slightly cooler, and therefore being unable to clearly favour either scale.

C11 used two different approaches, the infrared flux method (IRFM) and colour relations (CLBR) to determine stellar parameters for GCS stars. The IRFM is considered the most fundamental temperature scale after interferometry, and IRFM temperatures agree well with interferometric ones for stars in common ($\approx$30~K). The use of secondary colour--temperature--metallicity relations, however, which is necessary to fully cover the GCS catalogue, introduces an additional level of complication, so that it is worth testing the IRFM and the CLBR temperatures of C11 separately.

\begin{figure}
  \centering 
  \includegraphics[scale=0.4]{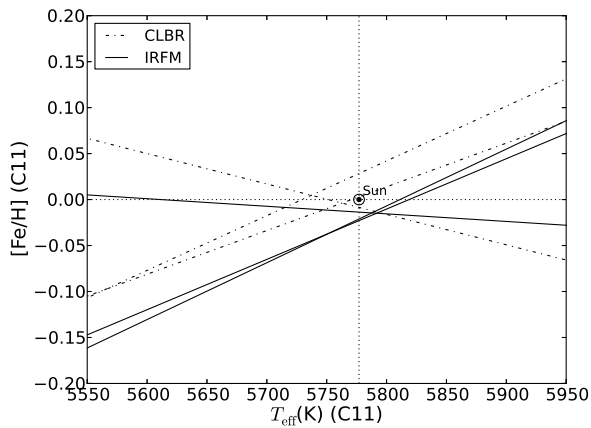}
  \caption{Degeneracy lines for the two subsamples in the C11. Stars that used IRFM calibrations, and stars that used CLBR calibrations. The lines of one type intersect at the point where the Sun would be in the sample. We also plot in comparison where the Sun really is, which is within the error of both samples. For clarity, we do not plot the errors.}
  \label{C11}
\end{figure}

In D14, using a sample of solar analogues from the HARPS archive, we have shown that overall we favour the temperature and metallicity scale in the C11 over the GCS-III. In this work, by taking advantage of our large combined FEplus sample, we examined whether the two different approaches in the C11 re--evaluation would result in different zero points for the temperature and metallicity scales. We divided the sample into two groups: those with IRFM parameters and those with CLBR parameters. These subsets consisted of 96 and 65 stars, respectively. We then applied our degeneracy lines method to determine the solar zero point for both samples (see Fig.\ref{C11} and Table~\ref{parameters}).

As can be seen from the figure, the zero points are the same within the errors, bracketing the position where the solar zero point should be. For the IRFM sample we derive $T_{\mathrm{eff}} = 5785\pm40$~K and [Fe/H] = $-0.02\pm0.02$~dex. For the CLBR sample we find $T_{\mathrm{eff}} = 5750\pm50$~K and [Fe/H] = $0.00\pm0.02$~dex. These values overlap within their errors and include the solar values. This shows that the subsets are mutually consistent, but the difference is similar to the uncertainty, leaving room for a difference in effective temperature of up to 35~K, which can be significant when comparing to the interferometric scale.

\section{Comparison of asteroids as solar reflectors}
\label{asteroids}

\begin{figure}
  \centering 
  \includegraphics[scale=0.4]{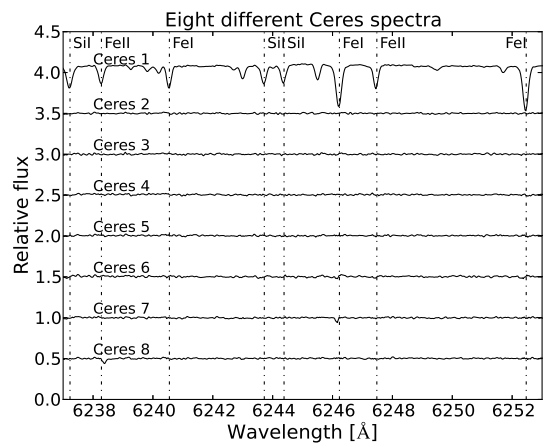}
  \caption{Small-wavelength window for all eight Ceres spectra taken during the December 2012 run (FE14 sample). It shows one spectrum and seven residual spectra, subtracted from the first.}
  \label{cerescomp}
\end{figure}

\begin{figure}
  \centering 
  \includegraphics[scale=0.4]{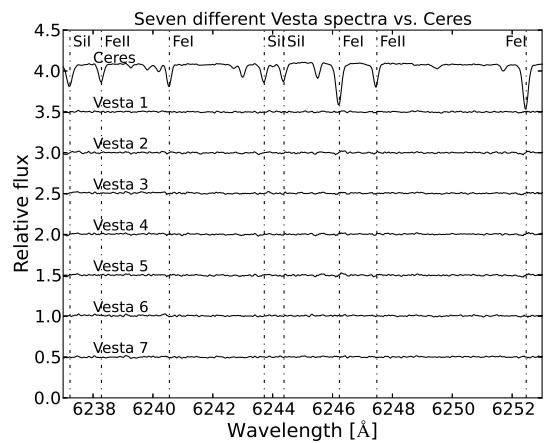}
  \caption{Same as Fig.~\ref{cerescomp}, but showing the residuals between Ceres and all seven Vesta spectra taken during the run.}
  \label{ceresvestacomp}
\end{figure}

To ensure that we have the best possible calibration from our solar comparison spectrum, taken from the asteroids Ceres and Vesta, we took spectra of them several times during the December 2012 run. In Fig.\ref{cerescomp} we show that the eight Ceres spectra do not change over the course of the run because the residuals show noise. In addition, we also plot the residuals between one Ceres and all Vesta spectra, see Fig.~\ref{ceresvestacomp}, showing that the choice of asteroid does not influence the analysis
either. Recently, \citet{Bd14} also showed in a very detailed analysis that the choice of asteroid and observation time, as long as it is during the night and not daytime \citep{Gr00}, does not influence the results of their very precise abundance analysis. In addition, \citet{Ki11} verified that the viewing angle of the reflection spectrum has no significant effect on the EW measurements, therefore we do not take that into account in this study.

\begin{figure}[h!]
\centering
\subfigure[Distribution of the quantities used in our first method for finding solar twins.]{
  \includegraphics[scale=0.35]{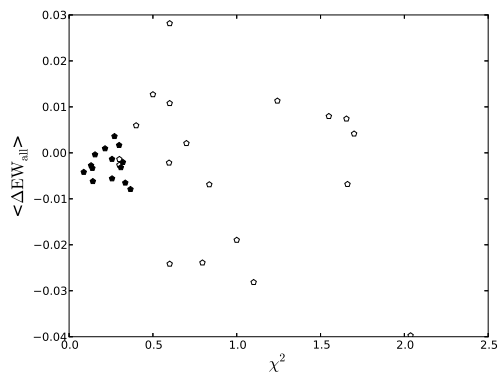}
\label{method1}
}
\subfigure[Distribution of the quantities used in our second method for finding solar twins.]{
  \includegraphics[scale=0.35]{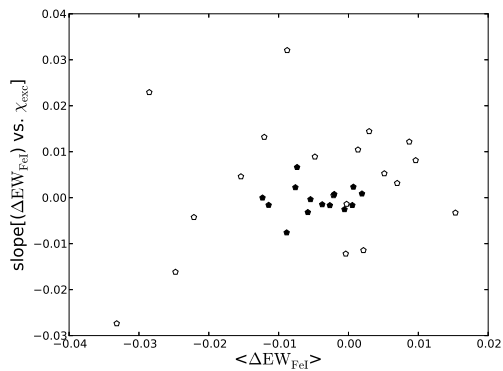}
\label{method2}
}
\subfigure[Distribution of the spectroscopic parameters determined for the asteroids and twins.]{
  \includegraphics[scale=0.35]{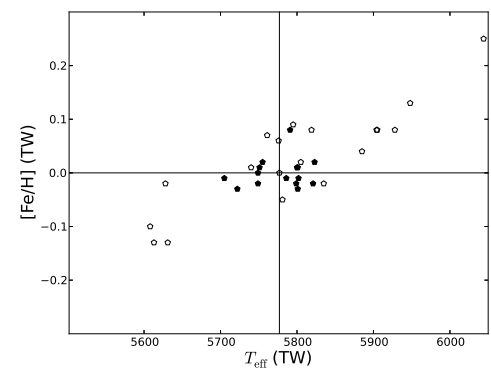}
\label{asttwins}
}
\caption[Optional caption for list of figures]{Panels (a) and (b) show the distribution of the main quantities used in the solar twin determination for the asteroids (filled pentagons) and the twins (empty pentagons). Clearly, the asteroids are more grouped together than the twins, and typically, only a few twins fall within the same area. Panel (c) shows spectroscopic $T_{\mathrm{eff}}$ and [Fe/H] values for the 15 asteroid spectra and the 19 solar twins from this work (TW). The twins that fall closest to the asteroid points should obviously be considered the closest twins in this study.}
\label{asteroidstwins}
\end{figure}

Following Sect.~\ref{solartwins}, we verified that the various reflected spectra would be seen as twins of each other, within our definition. The resulting values are $\chi^{2}(\Delta \mathrm{EW}_{\mathrm{all}}) \leq 0.37$, the median $\left<\Delta \mathrm{EW}_{\mathrm{all}}\right> \leq 0.008$, the median $\left<\Delta \mathrm{EW}_{\mathrm{FeI}}\right> \leq 0.01,$ and the slope[($\Delta \mathrm{EW}_{\mathrm{FeI}}$) vs. $\chi_{\mathrm{exc}}$] $\leq 0.008$ (Fig.~\ref{method1} and \ref{method2}). The figure shows how close to each other the asteroid spectra are (filled pentagons), in the differential analysis, as opposed to solar twins (open pentagons). Any spectrum falling within the ``asteroid cloud'' can be considered indistinguishable from the Sun according to that criterion (method i or ii). As expected, the solar twins are more scattered in the plot than the repeated solar (asteroid) spectra, yet there are twins that fall into the ``perfect twin'' region (although they are not necessarily the same twins in both plots).

In Fig.~\ref{asttwins} we show the spectroscopic parameters we determined with {\sc moog} and {\sc smh} for the asteroid spectra: although they are virtually indistinguishable in the differential comparison, the formal spectroscopic results for the asteroids spread from 5700 to 5830~K in effective temperature, but are typically better than $\pm0.03$~dex in metallicity (one exception). This agrees with the typical errors estimated in Sect.~\ref{specparam} from repeated stellar targets.

The solar twins cover an even broader range of values, and there clearly is a temperature--metallicity degeneracy still present in the model-dependent {\sc moog} analysis. The nine twins that fall within the area of the asteroids (roughly $\pm100$~K and $\pm0.1$~dex) should be considered especially close to being solar; they are marked with an asterisk in Table~\ref{twins}. They are HD28068, HD39769, HD76440, HD77461, HD78660, HD97356, HD138573, HD146233, and HD163441. The range $\pm100$~K and $\pm0.1$~dex from solar was also selected to be the solar twin range by \citet{Ra09},
but not all stars with these formal spectroscopic limits are to be considered twins (there would be 42 of them in our global FEplus sample), but only those that match the test of the differential comparison.

The twins with significantly different spectroscopic parameters typically lie within $\pm200$~K and $\pm0.15$~dex, with a clear temperature--metallicity degeneracy; therefore more accurate, dedicated differential line-by-line measurements \citep[e.g.][]{Me14,Bd14} will be necessary to determine whether this is an artefact of the automated, iterative model-dependent {\sc moog} analysis, or whether they are really not close twins. The most notable outlier is HD173071 ($T_{\mathrm{eff}} = 6044$~K and [Fe/H] = 0.25).

\section{Conclusions}

We have extended our previous study of stellar parameters for solar analogue stars. We examined two samples of FEROS spectra and solar comparison spectra (the asteroids Ceres and Vesta) and derived new spectroscopic metallicities, effective temperatures, and surface gravities for 148 solar analogue stars; 76 of them have previous spectroscopic values in the literature (in very good agreement with ours), while the rest are new determinations. Our degeneracy lines method (D12 and D14) shows that our spectroscopic parameters are well calibrated around the solar values, giving a solar zero point at $5755\pm40$~K and $0.00\pm0.02$~dex. Our values of effective temperature and metallicity are typically hotter and more metal rich by +65~K and +0.10~dex than those in the GCS-III catalogue from which our sample stars were selected.

We also showed that we can still find new solar twins, as given by our definition in Sect.~\ref{solartwins}. In D12 we had identified ten solar twins from the
first FEROS sample (six new to the literature). In the new sample of FEROS stars we find seven stars that we would consider to be solar twins: HD25680, HD29601, HD35769, HD39649, HD41708, HD67010, and HD76440. Of these seven, three have not been published as solar twins before this work. They are HD29601, HD39649, and HD67010. The star HD76440 was also part of the HARPS archive sample we analysed in D14 and the only twin in the FE14 sample we had identified before this work.

Using only the seventeen twins from both samples, we derive an average effective temperature and metallicity of $5799\pm128$~K and $0.03\pm0.10$~dex, again well centred on the Sun, even though with a large scatter.

Surface gravity varies for the stars in our sample between $\log g = 3.81 - 4.90$. However, considering only the stars we consider solar twins, their average surface gravity and scatter results in $4.45\pm0.10$, which agrees with the solar value of $4.44$.

Using the increased sample of 148 stars in the FEplus sample, we applied our degeneracy lines method and confirmed the offsets in the temperature and metallicity scale in the GCS-III around solar values, highlighted in D12 and D14, and found them to be present for the larger FEplus sample as well.

We then also investigated whether the line list used for our analysis might cause systematic differences. For this we chose four different lists from the literature and also combined them all into a comprehensive list. We then used these in our solar twin search and in several applications of our degeneracy line method. The results show that while there are some differences in the solar twins that the different lists select (but many of them are confirmed in many lists), there is no significant difference in the outcome of our degeneracy line method, which seems unaffected by the exact choice of spectral lines as long as they are weak, unblended, isolated lines on the linear part of the curve of growth.

We also further tested the C11 calibration by dividing our sample into subsamples, depending on how the stellar parameters were determined in C11, whether with the infrared-flux method (IRFM) or secondary colour calibrations (CLBR). We applied our degeneracy lines method to the two resulting subsamples and found the solar zero points to be at $5785\pm40$~K and $-0.02\pm0.02$~dex for the IRFM and $5750\pm50$~K and $0.00\pm0.02$~dex for the CLBR. These values are consistent with their errors. However, there is room for a difference of 35~K in effective temperature and 0.02~dex in metallicity, which adds to the problem of the current uncertainty on the absolute zero point of the temperature scale for large catalogues \citep{Ca14}.

\begin{acknowledgements}

We thank our referee Gustavo F. Porto de Mello for constructive remarks that helped improve the paper. We would also like to thank Jorge Mel{\'e}ndez and Andreas Korn for useful discussions. This study was financed by the Academy of Finland (grant nr.~130951 and 218317). We thank Swinburne University, where part of this work was carried out. 

\end{acknowledgements}


\onecolumn

\end{document}